\documentclass[onecolumn,secnumarabic,amssymb, nobibnotes, aps, prd]{revtex4-1}

\newcommand{\m}[1]{\macro{#1}}

\usepackage{graphicx}  
\usepackage{dcolumn}   
\usepackage{bm}        
\usepackage[english]{babel}
\usepackage{amsfonts,amsmath,amssymb,mathrsfs}
\usepackage{subfigure}
\usepackage{color}
\usepackage{hyperref}
\usepackage[left=2.5cm,top=2.6cm]{geometry}

\usepackage{ulem} 

\def\a{\alpha}
\def\at{\lambda}
\def\r{\rho}
\def\s{\sigma}
\def\t{\tau}
\def\m{\mu}
\def\n{\nu}
\def\k{\kappa}
\def\th{\theta}
\def\g{\gamma}\def\G{\Gamma}
\def\L{\Lambda}\def\l{\lambda}
\def\del{\nabla}
\def\D{\Delta}
\def\la{\langle}
\def\ra{\rangle}
\def\o{\omega}\def\O{\Omega}
\def\d{\delta}
\def\p{\partial}

\def\half{\textstyle{\frac{1}{2}}}

\def\bdoc{\begin{document}}
\def\edoc{\end{document}}

\def\beq{\begin{equation}}
\def\eeq{\end{equation}}
\def\bea{\begin{eqnarray}}
\def\eea{\end{eqnarray}}
\def\ben{\begin{enumerate}}
\def\een{\end{enumerate}}
\def\la{\langle}
\def\ra{\rangle}
\def\a{\alpha}
\def\b{\beta}
\def\g{\gamma}\def\G{\Gamma}
\def\d{\delta}\def\D{\Delta}
\def\e{\epsilon}
\def\z{\zeta}
\def \f {\frac}
\def\th{\theta}
\def\k{\kappa}
\def\l{\lambda}
\def\m{\mu}
\def\n{\nu}
\def\o{\omega}
\def\p{\pi}
\def\r{\rho}
\def\s{\sigma}
\def\t{\tau}
\def\L{${\cal L}$}
\def\H{${\cal H}$}
\def\S{\Sigma }
\def\gsim{\; \raisebox{-.8ex}{$\stackrel{\textstyle >}{\sim}$}\;}
\def\lsim{\; \raisebox{-.8ex}{$\stackrel{\textstyle <}{\sim}$}\;}
\def\gtrsim{\gsim}
\def\lessim{\lsim}
\def\loc{{\rm local}}
\def\vm{v_{\rm max}}
\def\bh{\bar{h}}
\def\del{\partial}
\def\nab{\nabla}
\def\half{{\textstyle{\frac{1}{2}}}}
\def\fourth{{\textstyle{\frac{1}{4}}}}

\def\bD{{\bf D}}
\def\bE{{\bf E}}
\def\bF{{\bf F}}
\def\bB{{\bf B}}
\def\bP{{\bf P}}
\def\bV{{\bf v}}
\def\bv{{\bf v}}
\def\bx{{\bf x}}
\def\by{{\bf y}}
\def\bz{{\bf z}}
\def\ba{{\bf a}}
\def\bd{{\bf d}}
\def\bs{{\bf s}}
\def\bn{{\bf n}}
\def\bp{{\bf p}}

\def\O{\Omega}

\def\del{\nabla}
\def\br{{\bf r}}
\def\bnab{{\bf \nab}}
\def\lf{\left (}
\def\rt{\right)}
\def\tE{\tilde{E}}
\def\tL{\tilde{L}}
\def\Horava{Ho\v{r}ava }
\def\le{\left }
\def\re{\right}

\def\ell{l}
\newcommand{\R}[4]{\ensuremath{R^{#1 #2}_{#3 #4}}}
\newcommand{\sgn}{{\rm sgn}}
\newcommand{\sD}[1]{\sum_{m=1}^{K}{#1}}
\newcommand{\Cal}[1]{\ensuremath{\mathcal{#1}}}

\begin{document}
\title{Membrane paradigm for Einstein-Gauss-Bonnet gravity}

\author{Ted Jacobson}
\email[email: ]{jacobson@umd.edu}
\affiliation{Maryland Center for Fundamental Physics, \\ University of Maryland, \\ College Park, Maryland 20742 USA}
\author{Arif Mohd}%
\email[email: ]{amohd@umd.edu}
\affiliation{Maryland Center for Fundamental Physics, \\ University of Maryland, \\ College Park, Maryland 20742 USA}
\author{Sudipta Sarkar}
\email[email: ]{sudiptas@iitgn.ac.in}
\affiliation{Indian Institute of Technology, \\ Gandhinagar, Gujarat, India 382355}

\date{\today}

\begin{abstract}
We construct the membrane paradigm for black objects in Einstein-Gauss-Bonnet gravity in spacetime dimensions $ \ge 5$. As in the case of general relativity, the horizon can be modeled as a membrane endowed with fluidlike properties. We derive the stress tensor for this membrane fluid and study the perturbation around static backgrounds with constant curvature horizon cross section, for which the stress tensor can be regularized with the usual redshift factor, and expressed in the form of a Newtonian viscous fluid with pressure, shear viscosity and bulk viscosity.
 We evaluate the transport coefficients for  black holes with constant curvature horizons and negative or zero cosmological constant. For the black brane geometry our result for the ratio of shear viscosity to entropy density agrees with that obtained previously in different frameworks.
\end{abstract}

\maketitle

\section{Introduction}
Gravitation, being the manifestation of the curvature of spacetime,
affects the causal structure of the spacetime. This can lead to the
existence of regions that are causally inaccessible to a class of
observers. An example of such a region is the portion of spacetime
inside the event horizon of a black hole, which is causally
disconnected from any outside observer.  Hence, the relevant physics
for the observers outside the hole must be independent of what is
happening inside the hole.  This observation forms the basis of the
membrane paradigm for black holes. \par
 In membrane paradigm
\cite{Thorne:1986iy, Damour:1978cg}  the
interaction of the black hole with the outside world is modeled by
replacing the black hole by a membrane of fictitious fluid ``living''
on the horizon.  Interaction of the black hole with the outside world
is then captured by the (theory dependent) transport coefficients of
the fluid. For example, the electromagnetic interaction of the black
hole is described by endowing the horizon with conductivity.  This
formalism provides an intuitive and elegant understanding of the
physics of the event horizon in terms of a simple nonrelativistic
language, and also serves as an efficient computational tool useful in
dealing with some astrophysical problems. After the advent of
holography, the membrane paradigm took a new life in which the
membrane fluid, at least in the case of planar horizons, captures some
aspects of the long wavelength description of the strongly coupled
quantum field theory at a finite temperature
\cite{Policastro:2001yc}. \par The original membrane paradigm
\cite{Thorne:1986iy, Damour:1978cg} was constructed for black holes in
general relativity.  The membrane fluid has the shear viscosity, $\eta
= 1/16 \pi G$. Dividing this by the Bekenstein-Hawking entropy
density, $ s = 1/4G$, gives a dimensionless number, $\eta/s = 1/4
\pi$. The calculation which leads to this ratio relies only on the
dynamics and the thermodynamics of the horizon in classical general
relativity. But, interestingly enough, it was found in
\cite{Policastro:2001yc} that the same ratio is obtained in the
holographic description of the hydrodynamic limit of the strongly
coupled $\mathcal{N}=4$, $U(N)$ gauge theory at finite temperature
which, is dual to general relativity in the limit of large $N$ and
large $\lambda_t$, where $N$ is the number of colors and
$\lambda_t$ is the `t Hooft coupling. The authors of
\cite{Kovtun:2003wp} conjectured that this ratio is a universal lower
bound for all the materials, this is called the KSS bound.  The
relationship between the membrane paradigm calculations and the
holographically derived KSS bound was explained as a consequence of
the trivial renormalization group (RG) flow from IR to UV in the
boundary gauge theory as one moves the outer cutoff surface from the
horizon to the boundary of spacetime
\cite{Iqbal:2008by,Bredberg:2010ky}. The universality of this bound,
how it might be violated, and the triviality of the RG flow in the
long wavelength limit at the level of the linear response, were first
clarified in \cite{Iqbal:2008by}. \par The general theory of
relativity, which is based upon the Einstein-Hilbert action
functional, is the simplest theory of gravity one can write guided by
the principle of diffeomorphism invariance while containing only the
time derivatives of second order in the equation of motion. Although
such a simple choice of the action functional has so far been adequate
to explain all the experimental and observational results, there is no
reason to believe that this choice is fundamental. Indeed, it is
expected on various general grounds that the low energy limit of any
quantum theory of gravity will contain higher derivative correction
terms. In fact, in string theory the low energy effective action
generically contains terms which are higher order in curvature due to
the stringy ($\alpha ')$ corrections. In the context of holographic
duality, such $\alpha '$ modifications correspond to the
corrections due to finite $\lambda_t$.  The specific form of these terms depends ultimately on
the detailed features of the quantum theory.  From the classical point
of view, a simple modification of the Einstein-Hilbert action is to
include the higher order curvature terms preserving the diffeomorphism
invariance and still leading to an equation of motion containing no
more than second order time derivatives. In fact this generalization
is unique \cite{Lovelock:1971yv} and goes by the name of
Lanczos-Lovelock gravity, of which the lowest order correction (second
order in curvature) appears as the Gauss-Bonnet (GB) term in spacetime
dimensions $D>4$. Einstein-Gauss-Bonnet (EGB) gravity is free from
ghosts \cite{Zwiebach:1985uq, Zumino:1985dp} (when linearized around
flat space), and leads to a
well-defined initial value problem. Black hole solutions in EGB
gravity have been studied extensively and are found to have various
interesting features \cite{Boulware:1985wk, Myers:1988ze}. The entropy
of these black holes is no longer proportional to the area of the
horizon but contains a curvature dependent term
\cite{Jacobson:1993xs,Wald:1993nt,Visser:1993nu,Iyer:1995kg}.  Hence,
unlike in general relativity, the entropy density of the horizon in
EGB gravity is not a constant but depends on the horizon
curvature. Now, the form of the membrane stress tensor in the fluid
model of the horizon is also theory dependent and therefore the
transport coefficients of the membrane fluid will change due to the
presence of the GB term in the action. Hence, it is of interest to
investigate the membrane paradigm and calculate the transport
coefficients for the membrane fluid in the EGB gravity. For planar
horizons the violation
of the KSS bound due to the GB term in the action has already been
shown in \cite{Brigante:2007nu,Brigante:2008gz} using other
methods. Also, the arguments that there really are string theories
that violate the bound were presented in \cite{Kats:2007mq,Buchel:2008vz}. \par
 In this paper, we extend to EGB gravity the action principle
formalism of the membrane paradigm as constructed in
\cite{Parikh:1997ma}. We first derive
the membrane stress-energy tensor on the stretched horizon for EGB
theory. After  restriction to the linearized
perturbations of static black backgrounds with horizon cross section
of constant curvature, followed by ``regularization,'' we express the membrane stress tensor in the
form of a Newtonian viscous fluid described by certain transport
coefficients. Our main result is the horizon curvature dependent membrane transport
coefficients in Eqs. \eqref{GB transport coeffBD}-\eqref{zeta},
and the ratio of 
shear viscosity and entropy density ($\eta/s$) in Eq. \eqref{ratio AdS black hole},
\beq
\frac{\eta}{s} =\frac{1}{4 \pi}
\frac{\left[1-2\lambda \frac{D-1}{D-3}\frac{1}{l^2}\right]+2 \lambda (D-5) \frac{k}{r_+^2}(1+\at \frac{k}{r_+^2})}{(1+2\at \frac{k}{r_+^2})
\left[1 +2 \at \frac{D-2}{D-4} \frac{k}{r_+^2}\right]}. 
\eeq
Here the horizon is a $D-2$ dimensional space of constant curvature
$(D-2)(D-3)k/r_+^2$ \, $(k=0,\pm 1)$, $\lambda = (D-3)(D-4) \alpha$,
 where $\alpha$ is the GB coupling
constant, and the cosmological constant is $\Lambda = -
(D-1)(D-2)/ 2 l^2$. The result for black holes in asymptotically flat spacetime follows by
taking the limit $l \to 0$.
In particular, for black brane $(k=0)$ in AdS, the ratio is \eqref{ratio for black brane} 
\beq \frac{\eta}{s} = \frac{1}{4\pi} \left[
1- 2\frac{(D-1)}{(D-3)} \frac{\lambda}{l^2} \right].
\eeq
 The presence of the GB term
violates the KSS bound, $\eta /s \geq 1/4\pi$, for any $\alpha >
0$. This matches with the result found in
\cite{Brigante:2007nu}, which used the  real-time AdS/CFT  calculation
of shear viscosity. Reference \cite{Brigante:2007nu} also applied the method of
\cite{Kovtun:2003wp} to calculate the ratio $\eta/s$ by relating it to  the 
 shear mode diffusion constant on the stretched horizon. The calculation
of the shear mode diffusion constant involves an integral that runs from the 
stretched horizon to infinity. In our membrane paradigm approach, by contrast, the
ratio is calculated at the horizon membrane, while the asymptotic spacetime plays
no role.

 This paper is organized as follows:  in Sec. \ref{geometric setup} we present the
 geometric setup of the membrane paradigm. In Sec. \ref{gr}, we review the action based
 membrane paradigm approach for the black holes in general relativity. This is then generalized in Sec. \ref{membrane GB} where we construct the membrane paradigm for the black objects in the EGB gravity. In that section we obtain the stress tensor for the membrane fluid and we derive the expressions for the transport coefficients of the fluid. In Sec. \ref{examples} we evaluate these transport coefficients for black holes with constant curvature horizons and negative or zero cosmological constant. Finally, we conclude with a summary and discussion in Sec. \ref{summary}. \par
 We adopt the metric signature $(-,+,+,+,...)$ and our sign
 conventions are the same as those of Misner-Thorne-Wheeler \cite{ Misner:1974qy}. All the symbols used in the main body of the paper are defined when introduced for the first time. For the convenience of the reader we have also included a table in the Appendix summarizing these symbols and their meanings.
\section{Geometric setup}
\label{geometric setup}
In this section we elaborate on the geometric setup necessary to construct the membrane paradigm. We include it in this paper to keep it self-contained but the interested reader can find a detailed discussion in the monograph \cite{Thorne:1986iy}. \par
The event horizon, $H$, of the black hole in $D$ spacetime dimensions, is a $(D\!-\!1)$-dimensional null hypersurface generated by the null geodesics $l^a$. We choose a nonaffine parametrization such that the null generators satisfy the geodesic equation $l^a \nabla_a l^b = \kappa \,l^b$, where $\kappa$ is a constant nonaffine coefficient. For a stationary spacetime, $l^a$ coincides
with the null limit of the timelike Killing vector and $\kappa$
can then be interpreted as the surface gravity of the horizon. \par
Next we introduce a timelike surface positioned just outside $H$ which is called the stretched horizon and denoted by $\mathcal{H}_s$. One can think of $\mathcal{H}_s$ as the world tube of a
family of fiducial observers just outside the black
hole horizon. The four velocity of these fiducial observers is denoted by $u^a$. Just
as $H$ is generated by the null congruence $l^a$, $\mathcal{H}_s$ is generated by the timelike congruence $u^a$. The unit normal to $\mathcal{H}_s$ is denoted by $n^a$ and is taken to point away from the horizon into the bulk.  We relate the points on $\mathcal{H}_s$ and $H$ by ingoing light rays parametrized by an affine parameter $\gamma$, such that $\gamma = 0$ is the position of the horizon and $(\partial / \partial \gamma)^a \, l_a = -1$ on the horizon. Then, in the limit $\gamma \to 0$, when the stretched horizon approaches the true one, $u^a \to \delta^{-1}l^a$
and $ n^a \to \delta^{-1} l^a$ where $\delta = \sqrt{2 \kappa \gamma}$ \cite{Thorne:1986iy}. 
If the black hole is stationary, and $l^a$ is the horizon generating Killing vector with surface gravity $\kappa$, then $\delta$ is the norm of the Killing vector, $\d = |l|$. 
\par
The induced metric $h_{ab}$ on $\mathcal{H}_s$ can be expressed in terms of the
spacetime metric $g_{ab}$ and the covariant normal  $n_a$ as $h_{ab} = g_{ab} - n_a n_b$. Similarly,
the induced metric $\gamma_{ab}$ on the $(D\!-\!2)$-dimensional 
spacelike cross section of $\mathcal{H}_s$ orthogonal to $u_a$ is given by $\gamma_{ab} =
h_{ab} + u_a u_b$. The extrinsic curvature of $\mathcal{H}_s$ is defined as $ K^{a}_{b} =
h^{c}_{b}\nabla_c n^a$. Using the limiting behavior of $n^a$ and $u^a$ it is easy to verify that in the limit that  $\delta \to 0$ various components of the extrinsic curvature behave as \cite{Thorne:1986iy}
\bea
\textrm{As $\delta \to 0$}&:& K^{u}_{u} = K^{a}_{b} u_a u^b = g  \sim \frac{\kappa}{\delta} ,\nonumber \\
&& K^{u}_{A} = K^a_b u_a \gamma^b_B = 0,\nonumber \\
&& K^{A}_{B} = K^a_b \gamma^A_a \gamma^b_B \sim \frac{k^{A}_{B}}{\delta}, \nonumber \\
&& K = K_{ab} g^{ab}\sim \frac{(\theta + \kappa)}{\delta} \label{nulllimit},
 \eea
where $\theta$ is the expansion scalar of $l^a$ and $k^A_B$ is the extrinsic curvature of the
$(D\!-\!2)$-dimensional spacelike cross section \footnote{$A,B$ denote the indices on the cross-section of the horizon.}of the true horizon $H$. Note that {\it a priori} the projection of the extrinsic curvature of $\mathcal{H}_s$ on the cross section of $\mathcal{H}_s$ has nothing to do with the extrinsic curvature of the cross section (orthogonal to $u^a$) {\it as embedded in} $\mathcal{H}_s$, i.e., there is, in general, no relationship between the pull-back of $\nabla_a n_b$ and $\nabla_a u_b$ to the cross section of the stretched horizon. However, in the null limit ($\delta \to 0 $) both $u^a$ and $n^a$ map to the same null vector $l^a$ and we have $K^A_B \to \delta^{-1} k^A_B$. Finally, we decompose
$k_{AB}$ into its trace-free and trace-full part as
\begin{eqnarray}
k_{AB} = \sigma_{AB} +  \frac{1}{(D-2)} \theta\, \gamma_{AB},
\end{eqnarray}
where $\sigma_{AB}$ is the shear of $l^a$. It is clear from Eq. \eqref{nulllimit} that in the
null limit, various components of the extrinsic curvature diverge
and we need to regularize them by multiplying by a factor of
$\delta$. The physical reason behind such infinities is that, as
the stretched horizon approaches the true one, the fiducial
observers experience more and more gravitational blueshift; on
the true horizon, the amount of blueshift is infinite. \par
This completes the description of our geometric setup. Next, we
review the derivation of the black hole membrane paradigm in
standard Einstein gravity.
\section{The membrane paradigm in Einstein gravity}
\label{gr} In this section we construct the membrane paradigm in
Einstein gravity in four spacetime dimensions. Our construction
will closely follow the action principle approach of \cite{Parikh:1997ma}. Our
purpose is to fix the notation and emphasize the points in the
construction which will be of importance for the corresponding
construction in the EGB gravity. We will highlight the
steps which will be different in the EGB case and where one has to make
assumptions. For the construction of the membrane stress tensor we will work exclusively with differential forms and
only in the end do we go back to the metric formalism. \par
In the rest of this paper, unless otherwise explicitly
written, we will work with the units such that  $16 \pi G = 1$.
The small Roman letters on the differential forms are the Lorentz indices while in the spacetime tensors we will not differentiate between the Lorentz and the world indices, this being understood that one can always use the vierbeins to convert the indices from the Lorentz to the world and vice versa. \par
 In the Cartan formalism, the Einstein-Hilbert Lagrangian is written in terms of the vector valued vierbein one-form $e^a$, related to the metric by
$g=\eta_{ab} e^a \otimes e^b$, and the Lorentz Lie-algebra valued torsion-free connection one-form $\omega^{ab}(e)$ defined by the equation $\mathcal{D}e^a=de^a + {\omega^a}_b \wedge e^b = 0$. The action is then 
\begin{equation}
S_{EH}=\frac{1}{2!}\int_M \Omega^{ab} \wedge e^c \wedge e^d
\epsilon_{abcd} \pm \frac{1}{2!}\int_{\partial M} \theta^{ab}
\wedge e^c \wedge e^d \epsilon_{abcd},
\end{equation}
where, $\Omega^{ab}$ is the curvature of
the torsion-free connection given by
$\Omega^{ab}=d\omega^{ab}+{\omega^a}_c \wedge \omega^{cb}$, and
$\theta^{ab}$ is the second fundamental form on the boundary
$\partial M$ of $M$, \cite{Myers:1987yn} \footnote
{Our sign convention is such that the plus (minus) sign applies to
the time-like (space-like) boundary.}. It is related to the extrinsic curvature by
\begin{equation}
\label{second form}
 \theta^{ab}=(n.n)(n^a K^b_c -n^b K^a_c)e^c.
\end{equation}
In our case, the boundary $\partial M$ of $M$ consists of the
outer boundary at spatial infinity and the inner boundary at the
stretched horizon, $\mathcal{H}_s$. Variation of the action with respect to $e^a$ can be separated into the contribution from $\omega(e)$ and the rest. Variation with respect to $\omega$ of the
bulk part of the action yields a total derivative which, after the
integration by parts, gives a contribution identical to the
negative of the variation of the boundary part. Thus the variation of the action
with respect to $\omega$ vanishes identically. In the absence of
the inner boundary, variation of $S_{EH}$ with respect to the
vierbein, $e^a$, under the Dirichlet boundary condition (holding
the vierbein fixed on the outer boundary) yields the equation of
motion for the vierbein. But when the inner boundary is present, there is no natural way to fix the vierbein there. The
physical reason for this is simply that the horizon acts as a
boundary only for a class of observers, and surely the metric is not fixed there. However, because it is a causal boundary the dynamics outside the horizon is not affected by what happens inside. Hence, for the consideration of the outside dynamics one can imagine some
fictitious matter living on the stretched horizon whose
contribution to the variation of the action cancels that of the
inner boundary. This is the basic idea of the construction of
stress-tensor {\it{\`{a} la}} Brown and York, \cite{Brown:1992br}.
Since we will be interested in the boundary term on the stretched
horizon which is a timelike hypersurface, from now on we will put
$n.n=1$. 

Hence, variation of the action with respect to the vierbein gives
\begin{equation}
\delta S_{EH} =\int_M \Omega^{ab} \wedge e^c \wedge \delta e^d
\epsilon_{abcd} + \int_{\partial M}\theta^{ab} \wedge e^c \wedge
\delta e^d \epsilon_{abcd}.
\end{equation}
The bulk term just gives the equation of motion. Using the
Dirichlet boundary condition on the outer boundary and the
expression of $\theta$ given in Eq. \eqref{second form}, the surviving contribution of the variation of the total action $S_{total}=S_{EH}+S_{matter}$ comes solely from the inner boundary and is given
by
\begin{align}
\delta S_{total} =  \int_{{\cal H}_s} K^b_m e^m \wedge e^c \wedge
\delta e^d \epsilon_{bcd},
\end{align}
where $\epsilon_{bcd} = - n^a
\epsilon_{abcd}$. 
This surviving contribution can be interpreted as due to a fictitious matter source residing on $\mathcal{H}_s$ whose stress tensor is given by 
\begin{equation}\label{membranetensorEH}
t^{ab}=2(K h^{ab}-K^{ab}).
\end{equation}
In terms of $t_{ab}$ the on-shell variation of $S_{EH}$ becomes
\begin{equation}
\delta S_{EH}=-\frac{1}{2}\int_{{\cal H}_s}t_{ab}\delta h^{ab}
+ \frac{1}{2} \int_M T_{ab} \delta g^{ab},
\end{equation}
where $T_{ab}=- \frac{2}{\sqrt{-g}}\frac{\delta
S_{matter}}{\delta g^{ab}}$ is the external matter's stress energy
tensor. Now consider the variation $\delta_{\xi}$ induced by a vector field $\xi$ which is arbitrary in the bulk and on the boundary behaves in a prescribed fashion.  We take $\xi $ to be such that it is  tangential to the inner boundary and vanishes on the all the other boundaries. Then the diffeomorphism invariance of the theory
ensures that 
\begin{align} 
\label{continuity}
\delta_{\xi} S_{EH} &= - \frac{1}{2} \int_{{\cal H}_s}t_{ab}\delta_{\xi} h^{ab} + \frac{1}{2} \int_M T_{ab} \delta_{\xi} g^{ab} = 0 \nonumber \\
&\Rightarrow - \int_{{\cal H}_s}t_{ab} D^a \xi^b +  \int_M T_{ab} \nabla^a \xi^b = 0,
 \end{align}
where, $D^a$ is the covariant derivative of the induced metric on the inner boundary $\mathcal{H}_s$, and $\nabla^a$ is the covariant derivative of the spacetime metric. Using integration by parts and the afore-mentioned conditions on $\xi^a$, Eq. \eqref{continuity} gives 
\beq \label {conservation EH} 
D^a t_{ab} = - T_{ac} n^a
{h^c}_b,
\eeq  
where the negative sign on the right-hand side arises because we have chosen $n^a$ as pointing away from the stretched horizon into the bulk. The right-hand
side of this equation can be interpreted as the flux of external
matter crossing the horizon from the bulk. Then Eq. \eqref{conservation EH} has the interpretation of the continuity
equation satisfied by the fictitious matter living on the stretched horizon.\par
At this stage, we would like to point out a difference between our approach and that of Ref. \cite{Parikh:1997ma}. In \cite{Parikh:1997ma}, the Gibbons-Hawking boundary term
 is considered to be only on the outer boundary. Therefore, one needs to show that a certain contribution containing the derivatives of the variation of the metric on the stretched horizon
 vanishes in the limit as the stretched horizon reaches the true horizon. In our approach there is no such requirement because we have the boundary term on the
 entire boundary which includes the stretched horizon. Since both  approaches finally yield the same horizon stress tensor, there is no difference in the physics in our approach and that of Ref. \cite{Parikh:1997ma}. However, an added advantage of our approach is that it also works beyond general relativity and, in particular, in any Lovelock gravity. If one does not include the Gibbons-Hawking term for Lovelock gravity, one needs to show that all the extra terms in the variation of the action vanish when the limit to the true horizon is taken. We have simply avoided this difficulty by adding the boundary term on the stretched horizon as well. Consequently, although our approach and that of Ref. \cite{Parikh:1997ma} ultimately give the same result for the horizon stress tensor, we believe the inclusion of the Gibbons-Hawking term makes the calculations easier. Also, it is worth pointing out that if we were to put the membrane at some finite distance from the horizon then we need the Hawking-Gibbons term in order to derive the stress tensor and the approach of Ref. \cite{Parikh:1997ma} just will not  work.
We believe that our approach is conceptually transparent and computationally simpler than the one in
 \cite{Parikh:1997ma}.
 \par
We have derived the form of the membrane stress tensor for the particular case of $D = 4$ spacetime dimensions, but it is easy to check that the form of the stress tensor in Eq. \eqref{membranetensorEH} remains unchanged for a general $D$-dimensional spacetime. Then the components of the membrane stress tensor $t_{ab}$, evaluated on the stretched horizon in the basis $(u^a,x^A)$ are given by
\bea
\label{membrane stress gr}
t_{uu}&=& \rho = -2\theta_{s}, \nonumber \\
t^{AB} &=&  2\left( -\sigma_{s}^{AB}+  \frac{(D-3)}{(D-2)}\theta_{s} \gamma^{AB}
+ g \,\gamma^{AB} \right),
\eea
where $ g = \kappa / \delta$. In deriving this expression, we have
replaced $K^{AB}$ by the expression  $\sigma_{s}^{AB}+
\frac{\theta_{s}}{(D-2)}\gamma^{AB}$ , where $\theta_{s}$ is the expansion and
$\sigma_{s}^{AB}$ is the shear of the
congruence generated by the timelike vector field $u^a$ on $\mathcal{H}_s$. As pointed out
in \cite{Thorne:1986iy,Damour:1978cg}, this replacement is valid only up to ${\cal
O}(\delta)$. Since we are ultimately interested in the limit
$\delta \to 0$, any ${\cal O}(\delta)$ error does not contribute.
Although this is certainly true for general relativity, for the EGB gravity such ${\cal O}(\delta)$ terms will play an important
role and they actually contribute in the limit that the stretched horizon
becomes the true horizon. \par
The particular form of the components of the
membrane stress tensor in Eq. \eqref{membrane stress gr} has an interpretation: the fictitious
matter on the stretched horizon can be regarded as a
$(D-2)$-dimensional viscous fluid \cite{Landau} with the energy density and 
transport coefficients given by 
\bea
\textrm {Energy density :}~&\rho_s = &-2 \theta_s , \nonumber \\
\textrm {Pressure : } ~ &p_s = &2g, \nonumber \\
\textrm {Shear viscosity : }~ &\eta_s = &1, \nonumber \\
\textrm {Bulk viscosity : }~ &\zeta_s = &- 2 \left(\frac{D-3}{D-2}\right). 
 \eea 
 Hence the entire $t_{ab}$ on the stretched horizon can be expressed as
 \bea \label{EHfluid}
t_{ab}=\rho_s u_a u_b +\gamma_a^A \gamma_b^B \left( p_s
\gamma_{AB} - 2 \eta_s {\sigma_s}_{AB} - \zeta_s \theta_s \gamma_{AB}
\right). 
\eea
Substituting these quantities in the conservation equation
\eqref{conservation EH} then gives the evolution equation for
the energy density, 
\bea \label{enEH} {\cal L}_u\rho_s + \rho_s
\theta_s = - p_s \theta_s + \zeta_s \theta_s^2 + 2 \eta_s
\sigma_s^2 + T_{ab} n^a u^b. 
\eea The evolution equation matches
exactly with the energy conservation equation of a viscous fluid. We
stress the fact that Eq. \eqref{enEH} is a direct
consequence of the conservation equation \eqref{conservation EH}
and the form of the stress tensor in Eq. \eqref{EHfluid}. In fact, in
any theory of gravity once we can express the stress tensor of the fictitious matter
obeying the conservation equation on the stretched horizon in a
form analogous to the one in Eq. \eqref{EHfluid}, the conservation equation will
automatically imply an evolution equation of the form
\eqref{enEH}. Notice that the conservation equation is only valid
on shell, which means that the equations of motion of the theory have
to be satisfied for it to hold.  \par
From the analysis of Sec. \ref{geometric setup} it is evident that as the stretched horizon
approaches the true horizon the membrane stress tensor in Eq. \eqref{membrane stress gr} diverges as $\delta^{-1}$ due to the large blueshift near the horizon. This divergence is regulated by simply
multiplying it by $\delta$. This
limiting and regularization procedure then yields a stress tensor
of a fluid living on the cross sections of the horizon itself in
terms of the quantities intrinsic to the horizon. This stress
tensor is
\bea
  t_{AB}^{(H)}=\left( p \gamma_{AB} - 2 \eta
\sigma_{AB} - \zeta \theta \gamma_{AB} \right),
\eea
and the energy density and the transport coefficients become,
 \bea
\textrm {Energy density :}~&\rho = &-2 \theta, \nonumber \\
\textrm {Pressure : } ~ &p = &2\kappa, \nonumber \\
\textrm {Shear viscosity : }~ &\eta = &1, \nonumber \\
\textrm {Bulk viscosity : }~ &\zeta = &- 2 \left(\frac{D-3}{D-2}\right)\eta, 
\label{GRcoefficients}
\eea
 where $\theta$ is the expansion of the
null generator of the true horizon as discussed in Sec. \ref{geometric setup}. Similarly, the regularization of the
evolution Eq. \eqref{enEH} gives \bea {\cal L}_l\rho + \rho
\theta = - p \theta + \zeta \theta^2 + 2 \eta {\sigma}^2 + T_{ab}
l^a l^b, \eea which is just the Raychaudhuri equation of the null
congruence generating the horizon. It should be noted that our approach  for deriving the evolution 
equation is different from the one used in \cite{Parikh:1997ma}. In principle,
one can just take the Lie derivative of the energy density with respect to the 
generator of the horizon to obtain the Raychaudhuri equation as in \cite{Parikh:1997ma} and then use the Einstein equation to replace the curvature dependence in terms of the matter energy-momentum tensor. We have followed an indirect approach in which we derive the evolution equation via the continuity equation \eqref{conservation EH} which is valid on shell, i.e., the equation of motion has already been used in its derivation. This approach is particularly useful when the equations of motion are complicated as in the EGB gravity and it becomes difficult to replace the curvature dependence in terms of the matter energy-momentum tensor. \par
For general relativity, the Bekenstein-Hawking entropy of the horizon is $ 4 \pi {\cal A} $, where ${\cal A}$ is the area of the horizon, hence the ratio of the shear viscosity to the entropy density is
\bea
\label{KSS bound}
\frac{\eta}{s} = \frac{1}{4\pi}.
\eea
Note that this is a dimensionless constant independent of the parameters of the horizon. Comparing this with the KSS bound, $\eta/s \geq 1/4\pi$, we see that the bound is saturated in general relativity. Interestingly, for any gravity theory with a Lagrangian depending on the Ricci scalar only, the value of this ratio is the same as that in general relativity and therefore the KSS bound is saturated in these theories \cite{Chatterjee:2010gp}. \par
Another important fact is that the bulk viscosity associated with the horizon is negative. Clearly, the fluid corresponding to the horizon is not an ordinary fluid, and as explained in  \cite{Thorne:1986iy}, {its negative bulk viscosity 
does not entail an instability, because the definition of the horizon imposes a future boundary condition precluding 
such behavior. 
\section{The membrane paradigm in Einstein-Gauss-Bonnet gravity}
\label{membrane GB}
EGB gravity is a natural generalization of
general relativity which includes terms higher order in 
curvature but in just such a way that the equation of motion
remains second order in time. \par The action of the theory is given
by \bea S_{total} = S_{EH} + \alpha S_{GB} + S_{matter} \eea where,
$S_{EH}$ and $S_{matter}$ are the contribution of the Einstein-Hilbert and
the matter, respectively, while $S_{GB}$ is the Gauss-Bonnet addition to the action. From the analysis in 
Sec. \ref{gr} we know how to take care of the $S_{EH}$. So, we can
exclusively work with the GB term now. The GB contribution to the
total action, in $D = 5$, is given by 
\footnote{For general $D \geq 4$, GB action is given by $S_{GB}= \int_M {\mathcal L}_m \pm \int_{\partial M} {\mathcal Q}_m$, where {\it m} is the greatest integer $\leq D/2$ and ${\mathcal L}_m = \Omega^{a_1 b_1} \wedge ... \wedge \Omega^{a_m b_m} \wedge \Sigma_{a_1 b_1 ... a_m b_m} $ and ${\mathcal Q}_m = m \int_0^1 \text{ds} ~ \theta^{a_1 b_1} \wedge \Omega_s^{a_2 b_2} 
\wedge ... \wedge \Omega_s^{a_m b_m} \wedge \Sigma_{a_1 b_1 ... a_m b_m}$. Here, $\Sigma_{a_1 b_1 \dots a_m b_m}= \frac{1}{(D-2m)!} \epsilon_{a_1b_1 \dots a_m b_m c_1\dots c_{D-2m}} e^{c_1} \wedge \dots \wedge e^{c_{D-2m}}$ and $\Omega_s$ is the curvature of the connection $\omega_s=\omega - s \theta$, $\Omega_s = d\omega_s + \omega_s \wedge \omega_s$. The sign of the surface term depends upon the timelike or spacelike nature of the boundary. The surface term was first constructed in Ref. \cite{Myers:1987yn} and more details can be found there.}
\cite{Myers:1987yn}
\bea
 \label{gb action} S_{GB}& = \int_M
\Omega^{ab} \wedge \Omega^{cd} \wedge e^f \epsilon_{abcdf} +
2\int_{\partial M} \theta^{ab} \wedge (\Omega-\frac{2}{3} \theta
\wedge \theta )^{cd}\wedge e^f \epsilon_{abcdf}.
\eea
As in the case of general relativity discussed in Sec. \ref{gr} the variation of
$S_{GB}$ with respect to the connection $\omega$ vanishes
identically. Variation with respect to the vierbien $e^a$ under the
Dirichlet boundary condition yields the equation of motion.
Using the torsion-free condition $\mathcal{D}e^a = 0$ on the connection,
this equation can be shown to be the same as that obtained in the
metric formalism. In the presence of the inner boundary at the
stretched horizon, $\mathcal{H}_s$,  we need the variation of the
boundary part of the  $S_{GB}$ due to the variation of the vierbein on
the inner boundary. This is obtained from Eq. \eqref{gb
action}, which, after variation with respect to $e^a$ and {\it then}
using the relations $\theta^{ab}=(n^a K^b_c - n^b
K^a_c)e^c $ and $\Omega^{ab}= \frac{1}{2} {R^{ab}}_{mn} e^m \wedge
e^n$, gives
\bea
 \delta S_{GB}|^{bndy}= 4 \int_{\partial M} K^s_a \left(
\frac{1}{2} {h^c_p h^d_q h_m^r h_n^s R^{pq}}_{rs} + \frac{2}{3} K^c_m K^d_n
\right) 4! \, \delta^{[amnb]}_{\,s\,c\,d\,f} \delta e^f_b.
 \eea 
The projections of the spacetime Riemann tensor can be written in terms of the Riemann tensor intrinsic to
$\mathcal{H}_s$ and the extrinsic curvature of $\mathcal{H}_s$
using the Gauss-Codazzi equation
 \bea \label{gauss-codazzi}
{h^c_p h^d_q h_a^r h_b^s R^{pq}}_{rs} = {\hat R^{cd}}{_{ab}} - K^c_a K^d_b + K^c_b K^d_a, \eea where ${\hat R^{cd}}{_{ab}}$ is the Riemann tensor intrinsic to $\mathcal{H}_s$. Thus
the variation of the boundary term can be written in terms of the
quantities intrinsic to the boundary,
 \bea
\delta S_{GB}|^{bndy}= 4 \int_{\partial M} K^s_a \left( \frac{1}{2}
{\hat R^{cd}}{_{m  n}} - \frac{1}{3} K^c_m K^d_n \right)
4! \, \delta^{[a mn b]}_{\,s\,c\,d\,f} \delta e^f_b.
\eea 
This can be evaluated to be 
\bea 
\label{delta S boundary GB}
\delta S_{GB}|^{bndy}= 4 \int_{\partial M} \left( 2 K_{mn} {\hat P^{a mn}}{_b} - J^a_b
+\frac{1}{3}J \delta^a_b \right) \delta e^b_a, 
\eea 
where we have defined,
\bea
{\hat P^{a mn}}{_b} &=&  {\hat R^{a mn}}{_b} + 2{\hat R^{n [ m}} h^{a]}_b + 2{\hat R^{[ a}}_b h^{m ] n } + \hat R h^{n [ a} h^{m]}_b , \\
J^a_b &=& K^2 K^a_b - K^{cd} K_{cd} K^a_b + 2 K^a_c K^c_d K^d_b - 2 K K^a_c K^c_b  ,\\
J &=& K^3-3 K  K^{cd} K_{cd}+2 K^a_b
K^b_c K^c_a. 
\eea
As in the case of general relativity, we can interpret the variation
$\delta S_{total}$ as due to a fictitious matter living on the
membrane whose stress energy tensor is given by 
the coefficient of $\delta e^a_b$. Thus from Eq. \eqref{delta S boundary GB} we can read off the membrane stress tensor due 
to the GB term (now including the GB coupling $\alpha$ and a negative sign arising from the fact that we are defining the stress tensor with covariant indices).
The total stress tensor for the
the membrane, including the contribution \eqref{membranetensorEH} coming from 
the Einstein-Hilbert action, is 
 \bea \label{total stress} t_{ab} =2 \left( K
h_{ab}-K_{ab} \right)- 4 \alpha \left( 2 K_{mn} {\hat P_a}{^{mn}}{_b} - J_{ab}
+\frac{1}{3}J h_{ab} \right). \eea
Although we have derived the membrane stress tensor for the particular case of $D = 5$, the result can be easily generalized to arbitrary dimensions and the form in Eq. \eqref{total stress} remains unchanged. By the same arguments as discussed in the case of general relativity, this $t_{ab}$ also satisfies the continuity equation \eqref{conservation EH}. This can also be verified explicitly by taking the divergence of $t_{ab}$ and using the appropriate projections of the equation of motion \cite{Davis:2002gn}. \par 
Note that a crucial difference between the membrane stress tensors for general relativity and EGB gravity is that in the former the stress tensor is linear in the extrinsic curvature while in the latter the stress tensor contains terms cubic in the extrinsic curvature of the stretched horizon.
 Since, as we take the limit
to the true horizon, the extrinsic curvature of the stretched horizon
diverges as $\delta^{-1}$, one would expect higher order
divergences in the case of EGB gravity coming from the contribution of the GB term to the stress tensor. The
regularization procedure used to tame the divergence coming from the Einstein-Hilbert term involves
multiplication with $\delta$, which does not tame the cubic order divergence coming from the GB term. Clearly one needs either a new
regularization procedure or some well-motivated
prescription which justifies neglect of the terms that lead to higher
order divergences in the limit when the stretched horizon
approaches the true one. In this paper, we will adopt the latter
approach. \par
We will restrict attention to 
background geometries which are static so that the expansion and the shear of the null generators of the
true horizon are zero.    Next, we will consider some arbitrary
perturbation of this
background which may arise due to the flux of matter flowing into
the horizon. As a result, the horizon becomes time dependent and
acquires expansion and shear. We will assume this perturbation of
the background geometry to be small so that we can work in the
linear order of perturbation and ignore all the higher order terms.
Essentially, our approximation mimics a slow physical-process
version of the dynamics of the horizon \cite{Jacobson:2003wv}. This essentially means that we are restricting
ourselves to the terms proportional to the first derivative
of the observer's four velocity $u^a$ which plays the role of the velocity
field for the fluid. We will discard all higher order
derivatives of the velocity except the linear one so that we can
write the membrane stress tensor as a Newtonian viscous fluid. In this limited setting we will see that
the only divergence that survives is of ${\cal O}{(\delta^{-1})}$
which can be
regularized in the same fashion as in the case of general relativity.
Therefore, when we encounter a product of two quantities $X$ and
$Y$, we will always express such a product as,
 \bea \label{perturbation scheme} 
 X Y \approx \mathring{X}\,  \mathring{Y} + \mathring{X} \,\delta Y +\mathring{Y} \, \delta X ,
  \eea 
  where $\mathring{X}$ is the value of the
quantity $X$ evaluated on the static background and $\delta X$ is the perturbed value of $X$ linear in perturbation. \par
In order to implement this scheme, we first define a quantity
$Q_{ab}$, whose importance will be apparent later, as \bea Q_{ab}
= K K_{ab} - K_{ac} K^c_b, \eea in terms of which, we write
 \bea
\label{factorized J}
J_{ab}=K_{ab} Q - 2 K_{ac} Q^c_b,
 \eea
 where $Q$ is the trace of $Q_{ab}$. \par
Now we observe the following facts. First, the components of the extrinsic curvature of $\mathcal{H}_s$ in the backgrounds that we are interested in is $\mathcal{O}(\delta)$. In particular, for the static spacetimes one can choose the cross sections of the $\mathcal{H}_s$ such that the pull-back of the extrinsic curvature to these cross sections is 
\beq\label{KAB}
K_{AB}=\frac{\delta}{r} \gamma_{AB}.
\eeq  
Here ``$r$" is defined by this equation; for the specific metrics considered below it will coincide with the 
radial coordinate in a particular coordinate system.
Second, for these backgrounds $Q_{ab}$ and $Q$ defined as above are finite on $\mathcal{H}_s$ and remain finite in the limit as $\mathcal{H}_s$ reaches the true horizon. In fact, $Q_{AB}$ for the background, in the limit that $\mathcal{H}_s$ approaches the true horizon, is simply $Q_{AB}=\frac{\kappa}{r} \gamma_{AB}$. Finally, the linearized $Q_{AB}$ and $Q$ have the most singular terms  given by
\bea
\label{linearlized Qs}
\delta Q_{AB}   &\sim &\frac{1}{\delta ^2} \kappa \left(\sigma_{AB} + \frac{\theta}{(D-2)} \gamma_{AB} \right), \\
\delta Q &\sim &   \frac{1}{\delta^2} 2 \kappa \theta.
\eea
Notice that the perturbation of the nonaffine coefficient $\kappa$ can always be gauged away by
choosing a suitable parametrization of the horizon. This remark means that we can put $\delta \kappa $ equal to zero. So, without any
loss of generality, we set $\kappa$ equal to the surface gravity
of the background black geometry. Now, as an illustration of the perturbation scheme mentioned in the Eq. \eqref{perturbation scheme} and the reason for the definition of the $Q_{ab}$, we use the facts mentioned in the previous paragraph to evaluate a term $K_{AC}Q^C_D$ contributed by $J_{AD}$ which we encounter in the projection of $t_{ab}$ on the cross section of $\mathcal{H}_s$. We evaluate this term as follows :
\bea
K_{AC}Q^C_D &\approx& \mathring{K}_{AC} \mathring{Q}^C_D + \delta K_{AC} \mathring{Q}^C_D+ \mathring{K}_{AC} \delta Q^C_D  \nonumber \\
&=& \frac{\delta}{r} \gamma_{AC} \mathring{Q}^C_D+ \frac{1}{\delta} \left(\sigma_{AC} + \frac{\theta}{(D-2)} \gamma_{AC} \right) \mathring{Q}^C_D + \frac{\delta}{r} \gamma_{AC} \frac{1}{\delta^2} \kappa 
\left( \sigma^C_D + \frac{\theta}{(D-2)} \gamma^C_D \right)  \nonumber \\
&\sim & \frac{1}{\delta} \left( \sigma_{AC} \mathring{Q}^C_D + \frac{\theta}{(D-2)} \mathring{Q}_{AD} + \frac{\kappa}{r} \sigma_{AD} + \frac{\kappa}{r} \frac{\theta}{(D-2)} \gamma_{AD} \right) \nonumber \\
&\sim & \frac{1}{\delta} \left( \frac{\kappa}{r}\sigma_{AD}  + \frac{\theta}{(D-2)} \frac{\kappa}{r} \gamma_{AD} + \frac{\kappa}{r} \sigma_{AD} + \frac{\kappa}{r} \frac{\theta}{(D-2)} \gamma_{AD} \right).
\eea
In the steps above we have dropped the terms which are of $\mathcal{O}(1)$ or $\mathcal{O}(\delta)$ because after regularization (i.e., multiplying by $\delta$) those terms will make no contribution. In this way one sees that our method of approximation is consistent. At the linear order in perturbation the only divergence that comes up in the membrane stress energy tensor is of $\mathcal{O}(\delta^{-1}) $ and therefore the whole stress tensor can be regularized simply by multiplying with $\delta$ exactly as in general relativity. The sample calculation above elaborates on how it is done for one particular term. Using the Gauss-Codazzi equation and the Raychaudhuri equation it can be shown that the projections of the curvature of the $\mathcal{H}_s$ on the cross section of $\mathcal{H}_s $ are equal to the curvature of the cross section in the limit $\delta \to 0$.  \par
Before writing down the stress tensor there is one more restriction that we are going to put on the background geometry. We require that the cross section of the horizon of the background geometry be a space of constant curvature, i.e., ${}^{(D-2)}\mathring{R}_{ABCD}=c \,(\gamma_{AC}\gamma_{BD}-\gamma_{AD}\gamma_{BC})$, where $c$ is a constant of dimension $length^{-2}$ which is related to the intrinsic Ricci scalar of the horizon cross section as
\bea
c = \frac{{\cal R}}{(D-3)(D-2)},\qquad {\cal R}\equiv {}^{(D-2)}\mathring{R}.
\eea
This assumption regarding the intrinsic geometry of the horizon cross section is necessary for the stress tensor to be of the form of an isotropic viscous fluid. Note that this assumption rules out, e.g., the topological
black holes in anti-de Sitter in which the horizon cross section is a nonconstant curvature Einstein space \cite{Birmingham:1998nr}. \par
 Using these observations and approximations, the contributions of different terms in the membrane stress energy tensor due to the GB term in the action are obtained as 
\begin{align}
\label{eq:limits}
K_{mn} {{\hat P}}_a{^{mn}}{_b} \gamma^a_A \gamma^b_B & \sim -\frac{1}{\delta} \frac{(D-4)}{2(D-2)}{\cal R} \kappa \gamma_{AB} + \frac{1}{\delta} \frac{(D-5)(D-4)}{2(D-3)(D-2)} {\cal R} \left[ \sigma_{AB} - \frac{(D-3)}{(D-2)} \theta \gamma_{AB} \right],\nonumber \\
J_{ab}\gamma^a_A \gamma^b_B &\sim \frac{1}{\delta} \sigma_{AB} (D-4) \frac{2 \kappa}{r} + \frac{1}{\delta} \gamma_{AB}  \frac{(D-3)}{(D-2)} \frac{4\kappa \theta}{r},  \nonumber \\
J &\sim \frac{1}{\delta} (D-3) \frac{6 \kappa \theta}{r} , \nonumber \\
K_{mn} {{\hat P}}_a{^{mn}}{_b} u^a u^b &\sim -\frac{1}{\delta} \frac{(D-4)}{2(D-2)} {\cal R} \kappa \theta,  \nonumber \\
J_{ab}u^a u^b &\sim  - \frac{1}{\delta} (D-3) \frac{2 \kappa \theta}{r}.
\end{align}
\par All the steps required to obtain the regularized membrane stress tensor are laid out now. Our perturbative strategy and the restriction that the horizon's cross section be the space of constant curvature  then yields the energy density and membrane stress tensor as 
\bea
\label{energy density}
\rho = -2 \theta \left[1 +  2\frac{D-4}{D-2} \alpha  {\cal R}\right], 
\eea
\bea
\label{fluid stress tensor}
t_{AB}^{(H)} = p\, \gamma_{AB} - 2\eta\, \sigma_{AB} - \zeta \theta\, \gamma_{AB},
\eea
where,
\begin{subequations}
 \label{GB transport coeff}
\begin{align}
p &= 2\kappa \left[1 + 2 \frac{D-4}{D-2}\alpha  {\cal R} \right], \\
\eta &=  1 - 4(D-4) \alpha \left[\frac{\kappa}{r} - \frac{(D-5)}{2(D-3)(D-2)} {\cal R}  \right],\\
\zeta &= -2 \frac{D-3}{D-2} \eta\label{relationetazeta}.
\end{align}
\end{subequations}
Note that the ratios $p/\rho=-\kappa/\theta$ and $\zeta/\eta= -2 \frac{D-3}{D-2}$ are the same as the ones that hold in general relativity \eqref{GRcoefficients}.  

We stress that while we are working only at the linear order in the perturbations, the $\alpha$ corrections in the transport coefficients given in Eq. \eqref{GB transport coeff} are nonperturbative. This simply means that the theory is exactly the EGB theory and the background spacetimes of interest are the exact static solutions of this theory. This approach differs from some of the other work in the literature, see for example \cite{Brustein:2008cg}, where one considers the effect of the GB term in the action as a small perturbation of general relativity. 

The continuity equation \eqref{conservation EH} yields the equation describing the evolution of the energy density along the null generators of the horizon. Note that we are actually applying the linearized continuity  equation, and in order to derive the evolution equation we have to keep the induced metric on the stretched horizon fixed. As in the case of general relativity, we first write down the equation on the stretched horizon, keeping the terms which are linear in perturbations and have $\mathcal{O}(\delta^{-1})$ divergence in $t_{ab}$, which gives an $\mathcal{O}(\delta^{-2})$ divergence in the continuity equation. This is regularized by multiplying the whole equation by $\delta^2$. This again has the same form as that in Eq. \eqref{enEH} with $\rho$ now given by \eqref{energy density}, and
$p$, $\eta$ and $\zeta$ given by \eqref{GB transport coeff}. 
\section{AdS black hole backgrounds and their various limits}
\label{examples}
In this section we will 
calculate the transport coefficients and the ratio $\eta / s$ for the membrane fluid
around EGB black hole backgrounds with a negative cosmological constant 
$\Lambda = - (D-1)(D-2)/2 l^2$ in the action, and horizon cross section of either 
positive, zero, or negative constant curvature.
The limit $\ell\rightarrow\infty$ will yield also asymptotically flat solutions.
The stable static solutions of this type in $D\ge5$ dimensions
is given by the metric \cite{Cai:2001dz}
\bea\label{ds2}
ds^2 = -f(r)dt^2 + \frac{1}{f(r)} dr^2 + r^2 h_{ij}dx^idx^j,
\eea 
where $h_{ij}$ is the metric of a $D-2$ dimensional space of constant curvature
$(D-2)(D-3)k$ and volume $\Sigma_k$, with $k=0,\pm 1$, and
\bea
f(r) = k+ \frac{r^2}{2  \at} \le[1-\sqrt{1+\frac{4\at M}{(D-2)\Sigma_k r^{D-1}} - \frac{4 \at}{l^2}}\re],
\eea
where 
\beq\label{alphatilde}
\at:=(D-3)(D-4)\a.
\eeq
The asymptotic AdS radius $L$ is defined by $\lim_{r\rightarrow\infty} f(r) =r^2/L^2$,
and is related to the cosmological 
constant length parameter $\ell$ in the Lagrangian by 
\beq
L^2/\ell^2 =\half \left(1+\sqrt{1-4\l/\ell^2}\right)
\eeq
The mass $M$ is related to the horizon radius $r_+$ by 
\beq\label{M}
M = (D-2)\Sigma_k r_+^{D-3}\left(k + \frac{\at k^2}{r_+^2} +\frac{r_+^2}{l^2}\right),
\eeq
the surface gravity is 
\beq\label{kappa}
\kappa= \frac{(D-1)r_+^4 + (D-3)kl^2r_+^2 +(D-5)\at k^2 l^2}{2 l^2 r_+(r_+^2 + 2\at k)},
\eeq
and the entropy density is 
\beq\label{s}
s= 4\pi\left(1+ \frac{(D-2)}{(D-4)}\frac{2\at k}{r_+^2}\right).
\eeq
We remind the reader that we are using units with $16\pi G=1$ \footnote{Our notation is different from the one in \cite{Brigante:2007nu}. We define the entropy density as the entropy per unit area of the horizon cross-section. For planar black holes the total entropy is infinite but the entropy density is well defined.}. Note that a hyperbolic ($k=-1$) horizon exists only if $\Lambda\ne0$ and/or $\at\ne0$, given the requirement that the mass \eqref{M} be positive.
The solution with $k=1$ was first found in Ref. \cite{Boulware:1985wk}. The cases $k=0$ and $-1$ were studied in Ref.\cite{Cai:2001dz}. Solutions for
particular value of the coupling, $\lambda = \ell^2/4$, were also studied in Refs. \cite{Cai:1998vy,Aros:2000ij,Banados:1993ur,Crisostomo:2000bb}.

Before we can evaluate the transport coefficients \eqref{GB transport coeff}, we must determine the parameter ``$r$" in the formula $K_{AB}=(\d/r)\gamma_{AB}$ \eqref{KAB} for the extrinsic curvature of the horizon cross section.
As explained in Sec.~\ref{geometric setup}, when the background is stationary with Killing vector $\xi$, we have $\d =|\xi|$. Consider then a metric of the form \eqref{ds2}, but with the coefficient $g_{rr}$ of $dr^2$ an independent function to begin with. The extrinsic curvature of the horizon in a $t=0$ slice is half the Lie derivative along the unit normal,
\beq
K_{AB}=\half{\cal L}_n \gamma_{AB} = \half \frac{d(r^2)}{ds} \frac{1}{r^{2}}\gamma_{AB}=  
\frac{1}{\sqrt{g_{rr}}}\frac{\gamma_{AB}}{r}.
\eeq
In the coordinate system of \eqref{ds2}, $1/\sqrt{g_{rr}}=\sqrt{g_{tt}}=\d$, so in fact ``$r$" in the formula for $K_{AB}$ is the same as the $r$ coordinate.

Using Eqs. \eqref{GB transport coeff}, \eqref{alphatilde}, and \eqref{kappa}, and ${\cal R}=(D-2)(D-3)k/r_+^2$, 
we find for the various transport coefficients 
\bea
 \label{GB transport coeffBD}
p &=&  \frac{(D-1)r_+}{l^2 } + \frac{(D-3)k}{r_+}+ \frac{(D-5)\at k^2}{r_+^{3}},  \\
\eta &=&  1 + \frac{2\at\bigl[-2\frac{k}{r_+^2}-(D-1)\frac{1}{l^2} +\lambda (D-5) (\frac{k}{r_+^2})^2\bigr]}{(D-3)(1+ 2 \at \frac{k}{r_+^2})}, \label{etahole} \\
\zeta &=& -2 \frac{(D-3)}{(D-2)} \,\eta. \label{zeta}
\eea
The ratio of shear viscosity and entropy density is given by
\beq
\label{ratio AdS black hole}
\frac{\eta}{s} =\frac{1}{4 \pi}
\frac{\left[1-2 \lambda \frac{D-1}{D-3}\frac{1}{l^2}\right]+2 \lambda (D-5) \frac{k}{r_+^2}(1+\at \frac{k}{r_+^2})}{(1+2\at \frac{k}{r_+^2})
\left[1 +2 \at \frac{D-2}{D-4} \frac{k}{r_+^2}\right]},
\eeq
which reduces to the GR value $1/4\pi$ when $\at=0$. Note that $r_+$ and $k$ always appear in the combination $k/r_+^2$ which is proportional to
the intrinsic Ricci scalar $\cal{R}$ of the horizon cross section. \par
In particular, for planar horizons k=0, and we get 
\bea
\label{ratio for black brane}
\frac{\eta}{s} = \frac{1}{4 \pi}\left[1-2 \frac{(D-1)}{(D-3)} \frac{\lambda}{l^2} \right],
\eea
which  matches with the value found by other methods in the literature, see for example \cite{Brigante:2007nu} where one of the calculations utilizes the Kaluza-Klein reduction to express the transverse metric perturbation in $D$-dimensions as the vector potential in $(D-1)$-dimensions and then uses the membrane paradigm results for the electromagnetic interaction of the black hole in $(D-1)$-dimensions. It is evident that the KSS bound is violated for any $\lambda > 0 $. The shear viscosity becomes negative when $\lambda/l^2  > (D-3)/2(D-1)$. It was pointed out in \cite{Brigante:2007nu,Iqbal:2008by} that the gravitons in the theory become strongly coupled as $\lambda/l^2 \to (D-3)/2(D-1)$, and for $\lambda/l^2  > (D-3)/2(D-1)$ the theory becomes unstable. 
\section{Summary and Discussion}
\label{summary}

In this paper we have employed a perturbative scheme to derive the horizon membrane stress tensor and the transport coefficients for the membrane fluid in Einstein-Gauss-Bonnet gravity.
We used the action principle formalism to determine the membrane stress tensor on the stretched horizon. Our derivation is slightly different from the one given in \cite{Parikh:1997ma}, 
since we include the Gibbons-Hawking boundary term at infinity as well as on the stretched horizon. As a result, the contribution from the bulk part under the variation with respect to $\omega$
 vanishes automatically. (In \cite{Parikh:1997ma} it was argued that, without the Gibbons-Hawking term, these contributions vanish in the limit when the stretched horizon approaches the true horizon.)  
Since the original posting of a draft of this paper, 
our method has been generalized to the higher order Lovelock theories \cite{Kolekar:2011gg}.\par
 
The membrane stress tensor in EGB gravity has terms cubic in the extrinsic curvature, and in the limit that the stretched horizon approaches the true horizon these terms are cubically divergent in $\delta^{-1}$. We avoided dealing with the cubic divergences by studying the perturbations about the background static black geometries. We found that restricting to the linear order in perturbations on the stretched horizon, the membrane stress tensor is in fact only linearly divergent. 
Therefore, at the linear order, the divergence structure of the GB contribution to the stress tensor is identical to that of the Einstein contribution. Hence the whole membrane stress tensor could be regularized in the same way as in general relativity: simply by multiplying the stress tensor by $\delta$.
 The source of the divergences can be traced to the fact that the time direction along the
stretched horizon becomes null in the limit $\d\rightarrow0$, so a finite passage of proper time
extends over an infinite affine parameter of the null geodesics generating the horizon, 
leading to divergent time derivatives.
These divergences could perhaps 
be avoided by working with the affine parameter directly on the true horizon, 
 rather than proper time on the stretched horizon If so, one could treat the nonlinear membrane dynamics without resorting to any perturbative scheme.\par 

In order to write the membrane stress tensor in the form of an isotropic viscous fluid we restricted the background geometries to those having a constant curvature horizon cross section. 
 It should be possible to lift this restriction by modeling the horizon as an anisotropic fluid with tensorial transport coefficients.  
The transport coefficients of the membrane fluid for the horizon in 
EGB gravity are given in Eq. \eqref{GB transport coeff}. 
These coefficients all receive contributions from the GB coupling, but we observe that 
the relation between the shear viscosity and bulk viscosity \eqref{relationetazeta}
is the same as it is in GR.
This also holds for the higher order Lovelock theories \cite{Kolekar:2011gg}, and for theories
whose Lagrangian is an arbitrary function of the Ricci scalar \cite{Chatterjee:2010gp}. 
In general relativity it is implied by the relative contributions of expansion and shear in the Raychaudhuri equation.
Given the assumption that the unperturbed horizon is maximally symmetric, perhaps the persistence of  
this relationship can be traced back to the structure of the Raychaudhuri equation in all these cases.
We also note that the ratio of the pressure to the energy density of the membrane fluid in EGB, and 
indeed in all Lovelock theories \cite{Kolekar:2011gg}, is the same as it is in GR. 

In Sec. \ref{examples} we evaluated the pressure, and bulk and shear viscosities for the perturbations of 
AdS black holes with flat or curved horizons. Results for black holes in flat space can be obtained by
taking the limit $l \to 0$.
 In particular, the ratio $\eta/s$ of shear viscosity $\eta$ to entropy density $s$ is given by Eq.~\eqref{ratio AdS black hole}.
For the black brane solution of EGB gravity, the membrane paradigm indeed gives a value for 
the ratio $\eta / s$
that agrees with  the literature. For example, in \cite{Brigante:2007nu}  the ratio is calculated using two different methods:
first, using a holographic calculation of the linear response function,
and second, using a Kaluza-Klein compactified version of the membrane paradigm. In agreement with that reference we find
that the KSS bound ($\eta/s = 1/4\pi$) is violated for any value of positive GB coupling.
It is worth mentioning that while our result for the ratio $\eta / s$ agrees with that of \cite{Buchel:2008vz}
our results for $\eta$ and $s$ separately do not agree. This can be attributed to the metric  field redefinition in \cite{Buchel:2008vz}, together with an inherent ambiguity in defining the spatial volume element of
the boundary gauge theory, which is a CFT and thus has no intrinsic scale. 
However, the dimensionless ratio $\eta/s$ is evidently invariant under these changes. 

In gauge/gravity duality, thermodynamic quantities pertaining to  black hole solutions in the gravity
theory, e.g., free energy, temperature,
and entropy etc., correspond to those describing the thermal state of the boundary gauge theory.
 It is therefore natural to expect that there would be some relation between the hydrodynamics 
 of the boundary theory and the long wavelength perturbations of the black hole solution.
At least for planar black hole
solutions, which have no length scale intrinsic to the horizon, one would expect a  hydrodynamic correspondence with the boundary gauge theory.  
Indeed, such a relation  was discovered, and is known as the fluid gravity correspondence \cite{Hubeny:2011hd}.  
Some transport coefficients describing the boundary fluid can also be calculated by the membrane paradigm.
 As shown in \cite{Iqbal:2008by},  when the bulk theory is general relativity, the low-frequency limit
 of the linear response of the boundary theory to certain perturbations (eg. shear, external current etc.) 
 can be calculated by that of the membrane-paradigm fluid living on the horizon of the black brane. 
The corresponding analysis when the bulk theory is  EGB gravity has not been carried out, 
as far as we are aware.

It is worth noting here that the negative bulk viscosity of
the horizon membrane fluid does not 
correspond to the value of the bulk viscosity for the boundary theory fluid.
The reason for this mismatch is that in the horizon membrane calculation
one is not really making a systematic expansion in derivatives of fluid velocity.
The study of the vacuum perturbation of a static black hole shows   
 \cite{Eling:2011ms} that the shear is second  order in perturbation, 
while the expansion is third order. Therefore,  while the shear viscosity of the boundary theory fluid
can be directly read from the shear term of the membrane stress tensor, the bulk viscosity of the 
boundary fluid is, in fact, zero. Perturbations due to matter, however, induce a lower order
nonzero expansion and yield a nonzero bulk viscosity \cite{Eling:2009pb}. 
The value of the bulk viscosity thus obtained 
has been shown, in general relativity, to agree with
the bulk viscosity of the boundary fluid \cite{Eling:2011ms,Buchel:2011yv}.

Finally, as mentioned above, we have found that, unlike in GR, the membrane transport coefficients in EGB gravity depend on the curvature of the horizon.  It is interesting to ask, then, under what circumstances, if any, this curvature dependence can tell us something about the transport coefficients in a dual CFT. For example, the membrane transport coefficients might provide a boundary condition for the flow equation determining the CFT transport coefficients, as discussed in \cite{{Iqbal:2008by}} (see also \cite{{Faulkner:2010jy}}). In order for the membrane fluid to be related to the hydrodynamic limit of the CFT,  we expect that the horizon curvature length scale must be much longer than the thermal length scale. This condition can be met for a large black hole ($r_+\gg\ell$). It can also be met when the  horizon curvature
is induced by the presence of an inhomogeneous background field, as in the setting of \cite{Donos:2017oym}.  It would be interesting to determine whether, under such circumstances, the membrane paradigm transport coefficients can indeed be related to those of a dual CFT.

\section*{Acknowledgments}
We would like to thank Don Marolf and Raman Sundrum for discussions, and Christopher Eling for useful correspondence. We are especially grateful to Robert Myers for 
detailed comments and helpful suggestions on a previous draft
of this article. A.M. would also like to thank Luca Bombelli for comments and discussions. S.S. thanks Maulik Parikh and Saugata Chatterjee for discussions during the initial phase of this work. This work was supported by NSF Grants No. PHY-0903572 and No. PHY-1407744.
\newpage
\section{Appendix: Table of important symbols and their meanings} 
\label{appendix}
$D$ denotes the spacetime dimension. The metric signature is $(-,+,+,...)$.
\begin{center}
  \begin{tabular}{|l | c | r | ll}
   \hline
   \bf{Symbol} & \bf{Meaning}  \\ \hline 
     & \\  $H$ & True horizon, a $(D-1)$-dimensional null-hypersurface  \\ \hline 
      & \\ $\mathcal{H}_s $  & Stretched horizon, a $(D-1)$-dimensional time-like hypersurface with \\ & tangent $u^a$ and normal $n^a$ \\  \hline
    & \\ $(a, b, c,...)$ & Spacetime indices\\  \hline
     &\\ $(A, B, C, ...)$ & Indices on the $(D-2)$-dimensional cross-section of the true/stretched horizon \\ \hline
     &\\ $l^a$ & Null generator of the true horizon parametrized by a non-affine parameter. \\ &Obeys the  geodesic equation: $l^a \nabla_a l^b = \kappa\, l^b$ \\ \hline
      &\\$h_{ab}$ & Induced metric on the stretched horizon \\ \hline
      &\\$\gamma_{ab}$ & Induced metric on the cross-section of the stretched horizon, which in the null limit is identified \\ & with the metric on the cross-section of the true horizon \\ \hline
      &\\$K_{ab}$ & Extrinsic curvature of the stretched horizon defined as, $ K_{ab} = \frac{1}{2} {\cal L}_{n} h_{ab}$ \\ \hline
      &\\$k_{AB}$ & Extrinsic curvature of the cross-section of the true horizon defined as $ k_{AB} = \frac{1}{2} {\cal L}_{l^a} \gamma_{AB}$\\ \hline
      &\\$\theta$, $\theta_s$ & Expansion of the true/strectched horizon \\ \hline
      &\\$\sigma^{ab}$, $\sigma_s^{ab}$ & Shear of the true/stretched horizon \\ \hline
     &\\$\hat{R}_{abcd}$ & Riemann tensor intrinsic to the stretched horizon \\ \hline
    & \\ $^{(D-2)}\mathring{R}_{ABCD}$ & Intrinsic Riemann tensor of the $(D-2)$-dimensional cross section of the \\ & stretched horizon in the background geometry, which in the null limit is identified \\ & with the intrinsic Riemann tensor of the cross-section of the true horizon \\ \hline
      &\\$^{(D-2)}\mathring{R}={\cal R}$ & Intrinsic Ricci scalar of the $(D-2)$-dimensional cross section of the \\ & stretched horizon in the background geometry, which in the null limit is identified \\ & with the intrinsic Ricci scalar of the cross-section of the true horizon \\ \hline
      &\\$\delta$& The parameter which measures the deviation of the stretched horizon from the true horizon\\
      & (equal to norm of horizon generating Killing vector when on a static background)\\ \hline
    & \\$c$&Defined as $c = \frac{^{(D-2)} \mathring{ R}}{(D-3)(D-2)}$ (For the planar horizon, $c = 0$) \\ \hline
    & \\$\alpha$ &Gauss-Bonnet coupling constant \\ \hline
     & \\$\lambda$ &Constant of dimension ${\it length}^2$ related to the Gauss-Bonnet coupling $\alpha$ as \\& $\lambda = (D-3)(D-4) \alpha$ \\ \hline
       \end{tabular}
\end{center}

\bibliography{membrane_GB}

\begin{thebibliography}{45}%
\makeatletter
\providecommand \@ifxundefined [1]{%
 \@ifx{#1\undefined}
}%
\providecommand \@ifnum [1]{%
 \ifnum #1\expandafter \@firstoftwo
 \else \expandafter \@secondoftwo
 \fi
}%
\providecommand \@ifx [1]{%
 \ifx #1\expandafter \@firstoftwo
 \else \expandafter \@secondoftwo
 \fi
}%
\providecommand \natexlab [1]{#1}%
\providecommand \enquote  [1]{``#1''}%
\providecommand \bibnamefont  [1]{#1}%
\providecommand \bibfnamefont [1]{#1}%
\providecommand \citenamefont [1]{#1}%
\providecommand \href@noop [0]{\@secondoftwo}%
\providecommand \href [0]{\begingroup \@sanitize@url \@href}%
\providecommand \@href[1]{\@@startlink{#1}\@@href}%
\providecommand \@@href[1]{\endgroup#1\@@endlink}%
\providecommand \@sanitize@url [0]{\catcode `\\12\catcode `\$12\catcode
  `\&12\catcode `\#12\catcode `\^12\catcode `\_12\catcode `\%12\relax}%
\providecommand \@@startlink[1]{}%
\providecommand \@@endlink[0]{}%
\providecommand \url  [0]{\begingroup\@sanitize@url \@url }%
\providecommand \@url [1]{\endgroup\@href {#1}{\urlprefix }}%
\providecommand \urlprefix  [0]{URL }%
\providecommand \Eprint [0]{\href }%
\providecommand \doibase [0]{http://dx.doi.org/}%
\providecommand \selectlanguage [0]{\@gobble}%
\providecommand \bibinfo  [0]{\@secondoftwo}%
\providecommand \bibfield  [0]{\@secondoftwo}%
\providecommand \translation [1]{[#1]}%
\providecommand \BibitemOpen [0]{}%
\providecommand \bibitemStop [0]{}%
\providecommand \bibitemNoStop [0]{.\EOS\space}%
\providecommand \EOS [0]{\spacefactor3000\relax}%
\providecommand \BibitemShut  [1]{\csname bibitem#1\endcsname}%
\let\auto@bib@innerbib\@empty
\bibitem [{\citenamefont {Thorne}\ \emph {et~al.}(1986)\citenamefont {Thorne},
  \citenamefont {Price},\ and\ \citenamefont {MacDonald}}]{Thorne:1986iy}%
  \BibitemOpen
  \bibfield  {author} {\bibinfo {author} {\bibfnamefont {K.}~\bibnamefont
  {Thorne}}, \bibinfo {author} {\bibfnamefont {R.}~\bibnamefont {Price}}, \
  and\ \bibinfo {author} {\bibfnamefont {D.}~\bibnamefont {MacDonald}},\ }\href
  {https://books.google.mu/books?id=T94hD5rR8oYC} {\emph {\bibinfo {title}
  {Black Holes: The Membrane Paradigm}}}\ (\bibinfo  {publisher} {Yale
  University Press},\ \bibinfo {year} {1986})\BibitemShut {NoStop}%
\bibitem [{\citenamefont {Damour}(1978)}]{Damour:1978cg}%
  \BibitemOpen
  \bibfield  {author} {\bibinfo {author} {\bibfnamefont {T.}~\bibnamefont
  {Damour}},\ }\href {\doibase 10.1103/PhysRevD.18.3598} {\bibfield  {journal}
  {\bibinfo  {journal} {Phys. Rev.}\ }\textbf {\bibinfo {volume} {D18}},\
  \bibinfo {pages} {3598} (\bibinfo {year} {1978})}\BibitemShut {NoStop}%
\bibitem [{\citenamefont {Policastro}\ \emph {et~al.}(2001)\citenamefont
  {Policastro}, \citenamefont {Son},\ and\ \citenamefont
  {Starinets}}]{Policastro:2001yc}%
  \BibitemOpen
  \bibfield  {author} {\bibinfo {author} {\bibfnamefont {G.}~\bibnamefont
  {Policastro}}, \bibinfo {author} {\bibfnamefont {D.~T.}\ \bibnamefont {Son}},
  \ and\ \bibinfo {author} {\bibfnamefont {A.~O.}\ \bibnamefont {Starinets}},\
  }\href {\doibase 10.1103/PhysRevLett.87.081601} {\bibfield  {journal}
  {\bibinfo  {journal} {Phys. Rev. Lett.}\ }\textbf {\bibinfo {volume} {87}},\
  \bibinfo {pages} {081601} (\bibinfo {year} {2001})},\ \Eprint
  {http://arxiv.org/abs/hep-th/0104066} {arXiv:hep-th/0104066 [hep-th]}
  \BibitemShut {NoStop}%
\bibitem [{\citenamefont {Kovtun}\ \emph {et~al.}(2003)\citenamefont {Kovtun},
  \citenamefont {Son},\ and\ \citenamefont {Starinets}}]{Kovtun:2003wp}%
  \BibitemOpen
  \bibfield  {author} {\bibinfo {author} {\bibfnamefont {P.}~\bibnamefont
  {Kovtun}}, \bibinfo {author} {\bibfnamefont {D.~T.}\ \bibnamefont {Son}}, \
  and\ \bibinfo {author} {\bibfnamefont {A.~O.}\ \bibnamefont {Starinets}},\
  }\href {\doibase 10.1088/1126-6708/2003/10/064} {\bibfield  {journal}
  {\bibinfo  {journal} {JHEP}\ }\textbf {\bibinfo {volume} {10}},\ \bibinfo
  {pages} {064} (\bibinfo {year} {2003})},\ \Eprint
  {http://arxiv.org/abs/hep-th/0309213} {arXiv:hep-th/0309213 [hep-th]}
  \BibitemShut {NoStop}%
\bibitem [{\citenamefont {Iqbal}\ and\ \citenamefont
  {Liu}(2009)}]{Iqbal:2008by}%
  \BibitemOpen
  \bibfield  {author} {\bibinfo {author} {\bibfnamefont {N.}~\bibnamefont
  {Iqbal}}\ and\ \bibinfo {author} {\bibfnamefont {H.}~\bibnamefont {Liu}},\
  }\href {\doibase 10.1103/PhysRevD.79.025023} {\bibfield  {journal} {\bibinfo
  {journal} {Phys. Rev.}\ }\textbf {\bibinfo {volume} {D79}},\ \bibinfo {pages}
  {025023} (\bibinfo {year} {2009})},\ \Eprint {http://arxiv.org/abs/0809.3808}
  {arXiv:0809.3808 [hep-th]} \BibitemShut {NoStop}%
\bibitem [{\citenamefont {Bredberg}\ \emph {et~al.}(2011)\citenamefont
  {Bredberg}, \citenamefont {Keeler}, \citenamefont {Lysov},\ and\
  \citenamefont {Strominger}}]{Bredberg:2010ky}%
  \BibitemOpen
  \bibfield  {author} {\bibinfo {author} {\bibfnamefont {I.}~\bibnamefont
  {Bredberg}}, \bibinfo {author} {\bibfnamefont {C.}~\bibnamefont {Keeler}},
  \bibinfo {author} {\bibfnamefont {V.}~\bibnamefont {Lysov}}, \ and\ \bibinfo
  {author} {\bibfnamefont {A.}~\bibnamefont {Strominger}},\ }\href {\doibase
  10.1007/JHEP03(2011)141} {\bibfield  {journal} {\bibinfo  {journal} {JHEP}\
  }\textbf {\bibinfo {volume} {03}},\ \bibinfo {pages} {141} (\bibinfo {year}
  {2011})},\ \Eprint {http://arxiv.org/abs/1006.1902} {arXiv:1006.1902
  [hep-th]} \BibitemShut {NoStop}%
\bibitem [{\citenamefont {Lovelock}(1971)}]{Lovelock:1971yv}%
  \BibitemOpen
  \bibfield  {author} {\bibinfo {author} {\bibfnamefont {D.}~\bibnamefont
  {Lovelock}},\ }\href {\doibase 10.1063/1.1665613} {\bibfield  {journal}
  {\bibinfo  {journal} {J. Math. Phys.}\ }\textbf {\bibinfo {volume} {12}},\
  \bibinfo {pages} {498} (\bibinfo {year} {1971})}\BibitemShut {NoStop}%
\bibitem [{\citenamefont {Zwiebach}(1985)}]{Zwiebach:1985uq}%
  \BibitemOpen
  \bibfield  {author} {\bibinfo {author} {\bibfnamefont {B.}~\bibnamefont
  {Zwiebach}},\ }\href {\doibase 10.1016/0370-2693(85)91616-8} {\bibfield
  {journal} {\bibinfo  {journal} {Phys. Lett.}\ }\textbf {\bibinfo {volume}
  {B156}},\ \bibinfo {pages} {315} (\bibinfo {year} {1985})}\BibitemShut
  {NoStop}%
\bibitem [{\citenamefont {Zumino}(1986)}]{Zumino:1985dp}%
  \BibitemOpen
  \bibfield  {author} {\bibinfo {author} {\bibfnamefont {B.}~\bibnamefont
  {Zumino}},\ }\href {\doibase 10.1016/0370-1573(86)90076-1} {\bibfield
  {journal} {\bibinfo  {journal} {Phys. Rept.}\ }\textbf {\bibinfo {volume}
  {137}},\ \bibinfo {pages} {109} (\bibinfo {year} {1986})}\BibitemShut
  {NoStop}%
\bibitem [{\citenamefont {Boulware}\ and\ \citenamefont
  {Deser}(1985)}]{Boulware:1985wk}%
  \BibitemOpen
  \bibfield  {author} {\bibinfo {author} {\bibfnamefont {D.~G.}\ \bibnamefont
  {Boulware}}\ and\ \bibinfo {author} {\bibfnamefont {S.}~\bibnamefont
  {Deser}},\ }\href {\doibase 10.1103/PhysRevLett.55.2656} {\bibfield
  {journal} {\bibinfo  {journal} {Phys. Rev. Lett.}\ }\textbf {\bibinfo
  {volume} {55}},\ \bibinfo {pages} {2656} (\bibinfo {year}
  {1985})}\BibitemShut {NoStop}%
\bibitem [{\citenamefont {Myers}\ and\ \citenamefont
  {Simon}(1988)}]{Myers:1988ze}%
  \BibitemOpen
  \bibfield  {author} {\bibinfo {author} {\bibfnamefont {R.~C.}\ \bibnamefont
  {Myers}}\ and\ \bibinfo {author} {\bibfnamefont {J.~Z.}\ \bibnamefont
  {Simon}},\ }\href {\doibase 10.1103/PhysRevD.38.2434} {\bibfield  {journal}
  {\bibinfo  {journal} {Phys. Rev.}\ }\textbf {\bibinfo {volume} {D38}},\
  \bibinfo {pages} {2434} (\bibinfo {year} {1988})}\BibitemShut {NoStop}%
\bibitem [{\citenamefont {Jacobson}\ and\ \citenamefont
  {Myers}(1993)}]{Jacobson:1993xs}%
  \BibitemOpen
  \bibfield  {author} {\bibinfo {author} {\bibfnamefont {T.}~\bibnamefont
  {Jacobson}}\ and\ \bibinfo {author} {\bibfnamefont {R.~C.}\ \bibnamefont
  {Myers}},\ }\href {\doibase 10.1103/PhysRevLett.70.3684} {\bibfield
  {journal} {\bibinfo  {journal} {Phys. Rev. Lett.}\ }\textbf {\bibinfo
  {volume} {70}},\ \bibinfo {pages} {3684} (\bibinfo {year} {1993})},\ \Eprint
  {http://arxiv.org/abs/hep-th/9305016} {arXiv:hep-th/9305016 [hep-th]}
  \BibitemShut {NoStop}%
\bibitem [{\citenamefont {Wald}(1993)}]{Wald:1993nt}%
  \BibitemOpen
  \bibfield  {author} {\bibinfo {author} {\bibfnamefont {R.~M.}\ \bibnamefont
  {Wald}},\ }\href {\doibase 10.1103/PhysRevD.48.R3427} {\bibfield  {journal}
  {\bibinfo  {journal} {Phys. Rev.}\ }\textbf {\bibinfo {volume} {D48}},\
  \bibinfo {pages} {R3427} (\bibinfo {year} {1993})},\ \Eprint
  {http://arxiv.org/abs/gr-qc/9307038} {arXiv:gr-qc/9307038 [gr-qc]}
  \BibitemShut {NoStop}%
\bibitem [{\citenamefont {Visser}(1993)}]{Visser:1993nu}%
  \BibitemOpen
  \bibfield  {author} {\bibinfo {author} {\bibfnamefont {M.}~\bibnamefont
  {Visser}},\ }\href {\doibase 10.1103/PhysRevD.48.5697} {\bibfield  {journal}
  {\bibinfo  {journal} {Phys. Rev.}\ }\textbf {\bibinfo {volume} {D48}},\
  \bibinfo {pages} {5697} (\bibinfo {year} {1993})},\ \Eprint
  {http://arxiv.org/abs/hep-th/9307194} {arXiv:hep-th/9307194 [hep-th]}
  \BibitemShut {NoStop}%
\bibitem [{\citenamefont {Iyer}\ and\ \citenamefont
  {Wald}(1995)}]{Iyer:1995kg}%
  \BibitemOpen
  \bibfield  {author} {\bibinfo {author} {\bibfnamefont {V.}~\bibnamefont
  {Iyer}}\ and\ \bibinfo {author} {\bibfnamefont {R.~M.}\ \bibnamefont
  {Wald}},\ }\href {\doibase 10.1103/PhysRevD.52.4430} {\bibfield  {journal}
  {\bibinfo  {journal} {Phys. Rev.}\ }\textbf {\bibinfo {volume} {D52}},\
  \bibinfo {pages} {4430} (\bibinfo {year} {1995})},\ \Eprint
  {http://arxiv.org/abs/gr-qc/9503052} {arXiv:gr-qc/9503052 [gr-qc]}
  \BibitemShut {NoStop}%
\bibitem [{\citenamefont {Brigante}\ \emph
  {et~al.}(2008{\natexlab{a}})\citenamefont {Brigante}, \citenamefont {Liu},
  \citenamefont {Myers}, \citenamefont {Shenker},\ and\ \citenamefont
  {Yaida}}]{Brigante:2007nu}%
  \BibitemOpen
  \bibfield  {author} {\bibinfo {author} {\bibfnamefont {M.}~\bibnamefont
  {Brigante}}, \bibinfo {author} {\bibfnamefont {H.}~\bibnamefont {Liu}},
  \bibinfo {author} {\bibfnamefont {R.~C.}\ \bibnamefont {Myers}}, \bibinfo
  {author} {\bibfnamefont {S.}~\bibnamefont {Shenker}}, \ and\ \bibinfo
  {author} {\bibfnamefont {S.}~\bibnamefont {Yaida}},\ }\href {\doibase
  10.1103/PhysRevD.77.126006} {\bibfield  {journal} {\bibinfo  {journal} {Phys.
  Rev.}\ }\textbf {\bibinfo {volume} {D77}},\ \bibinfo {pages} {126006}
  (\bibinfo {year} {2008}{\natexlab{a}})},\ \Eprint
  {http://arxiv.org/abs/0712.0805} {arXiv:0712.0805 [hep-th]} \BibitemShut
  {NoStop}%
\bibitem [{\citenamefont {Brigante}\ \emph
  {et~al.}(2008{\natexlab{b}})\citenamefont {Brigante}, \citenamefont {Liu},
  \citenamefont {Myers}, \citenamefont {Shenker},\ and\ \citenamefont
  {Yaida}}]{Brigante:2008gz}%
  \BibitemOpen
  \bibfield  {author} {\bibinfo {author} {\bibfnamefont {M.}~\bibnamefont
  {Brigante}}, \bibinfo {author} {\bibfnamefont {H.}~\bibnamefont {Liu}},
  \bibinfo {author} {\bibfnamefont {R.~C.}\ \bibnamefont {Myers}}, \bibinfo
  {author} {\bibfnamefont {S.}~\bibnamefont {Shenker}}, \ and\ \bibinfo
  {author} {\bibfnamefont {S.}~\bibnamefont {Yaida}},\ }\href {\doibase
  10.1103/PhysRevLett.100.191601} {\bibfield  {journal} {\bibinfo  {journal}
  {Phys. Rev. Lett.}\ }\textbf {\bibinfo {volume} {100}},\ \bibinfo {pages}
  {191601} (\bibinfo {year} {2008}{\natexlab{b}})},\ \Eprint
  {http://arxiv.org/abs/0802.3318} {arXiv:0802.3318 [hep-th]} \BibitemShut
  {NoStop}%
\bibitem [{\citenamefont {Kats}\ and\ \citenamefont
  {Petrov}(2009)}]{Kats:2007mq}%
  \BibitemOpen
  \bibfield  {author} {\bibinfo {author} {\bibfnamefont {Y.}~\bibnamefont
  {Kats}}\ and\ \bibinfo {author} {\bibfnamefont {P.}~\bibnamefont {Petrov}},\
  }\href {\doibase 10.1088/1126-6708/2009/01/044} {\bibfield  {journal}
  {\bibinfo  {journal} {JHEP}\ }\textbf {\bibinfo {volume} {01}},\ \bibinfo
  {pages} {044} (\bibinfo {year} {2009})},\ \Eprint
  {http://arxiv.org/abs/0712.0743} {arXiv:0712.0743 [hep-th]} \BibitemShut
  {NoStop}%
\bibitem [{\citenamefont {Buchel}\ \emph {et~al.}(2009)\citenamefont {Buchel},
  \citenamefont {Myers},\ and\ \citenamefont {Sinha}}]{Buchel:2008vz}%
  \BibitemOpen
  \bibfield  {author} {\bibinfo {author} {\bibfnamefont {A.}~\bibnamefont
  {Buchel}}, \bibinfo {author} {\bibfnamefont {R.~C.}\ \bibnamefont {Myers}}, \
  and\ \bibinfo {author} {\bibfnamefont {A.}~\bibnamefont {Sinha}},\ }\href
  {\doibase 10.1088/1126-6708/2009/03/084} {\bibfield  {journal} {\bibinfo
  {journal} {JHEP}\ }\textbf {\bibinfo {volume} {03}},\ \bibinfo {pages} {084}
  (\bibinfo {year} {2009})},\ \Eprint {http://arxiv.org/abs/0812.2521}
  {arXiv:0812.2521 [hep-th]} \BibitemShut {NoStop}%
\bibitem [{\citenamefont {Parikh}\ and\ \citenamefont
  {Wilczek}(1998)}]{Parikh:1997ma}%
  \BibitemOpen
  \bibfield  {author} {\bibinfo {author} {\bibfnamefont {M.}~\bibnamefont
  {Parikh}}\ and\ \bibinfo {author} {\bibfnamefont {F.}~\bibnamefont
  {Wilczek}},\ }\href {\doibase 10.1103/PhysRevD.58.064011} {\bibfield
  {journal} {\bibinfo  {journal} {Phys. Rev.}\ }\textbf {\bibinfo {volume}
  {D58}},\ \bibinfo {pages} {064011} (\bibinfo {year} {1998})},\ \Eprint
  {http://arxiv.org/abs/gr-qc/9712077} {arXiv:gr-qc/9712077 [gr-qc]}
  \BibitemShut {NoStop}%
\bibitem [{\citenamefont {Misner}\ \emph {et~al.}(1973)\citenamefont {Misner},
  \citenamefont {Thorne},\ and\ \citenamefont {Wheeler}}]{Misner:1974qy}%
  \BibitemOpen
  \bibfield  {author} {\bibinfo {author} {\bibfnamefont {C.~W.}\ \bibnamefont
  {Misner}}, \bibinfo {author} {\bibfnamefont {K.~S.}\ \bibnamefont {Thorne}},
  \ and\ \bibinfo {author} {\bibfnamefont {J.~A.}\ \bibnamefont {Wheeler}},\
  }\href@noop {} {\emph {\bibinfo {title} {{Gravitation}}}}\ (\bibinfo
  {publisher} {W. H. Freeman},\ \bibinfo {address} {San Francisco},\ \bibinfo
  {year} {1973})\BibitemShut {NoStop}%
\bibitem [{Note1()}]{Note1}%
  \BibitemOpen
  \bibinfo {note} {$A,B$ denote the indices on the cross-section of the
  horizon.}\BibitemShut {Stop}%
\bibitem [{\citenamefont {Myers}(1987)}]{Myers:1987yn}%
  \BibitemOpen
  \bibfield  {author} {\bibinfo {author} {\bibfnamefont {R.~C.}\ \bibnamefont
  {Myers}},\ }\href {\doibase 10.1103/PhysRevD.36.392} {\bibfield  {journal}
  {\bibinfo  {journal} {Phys. Rev.}\ }\textbf {\bibinfo {volume} {D36}},\
  \bibinfo {pages} {392} (\bibinfo {year} {1987})}\BibitemShut {NoStop}%
\bibitem [{Note2()}]{Note2}%
  \BibitemOpen
  \bibinfo {note} {Our sign convention is such that the plus (minus) sign
  applies to the time-like (space-like) boundary.}\BibitemShut {Stop}%
\bibitem [{\citenamefont {Brown}\ and\ \citenamefont
  {York}(1993)}]{Brown:1992br}%
  \BibitemOpen
  \bibfield  {author} {\bibinfo {author} {\bibfnamefont {J.~D.}\ \bibnamefont
  {Brown}}\ and\ \bibinfo {author} {\bibfnamefont {J.~W.}\ \bibnamefont {York},
  \bibfnamefont {Jr.}},\ }\href {\doibase 10.1103/PhysRevD.47.1407} {\bibfield
  {journal} {\bibinfo  {journal} {Phys. Rev.}\ }\textbf {\bibinfo {volume}
  {D47}},\ \bibinfo {pages} {1407} (\bibinfo {year} {1993})},\ \Eprint
  {http://arxiv.org/abs/gr-qc/9209012} {arXiv:gr-qc/9209012 [gr-qc]}
  \BibitemShut {NoStop}%
\bibitem [{\citenamefont {Landau}\ and\ \citenamefont
  {Lifshitz}(1987)}]{Landau}%
  \BibitemOpen
  \bibfield  {author} {\bibinfo {author} {\bibfnamefont {L.~D.}\ \bibnamefont
  {Landau}}\ and\ \bibinfo {author} {\bibfnamefont {E.~M.}\ \bibnamefont
  {Lifshitz}},\ }\href@noop {} {\emph {\bibinfo {title} {{Fluid Mechanics,
  Second Edition: Volume 6 (Course of Theoretical Physics)}}}}\ (\bibinfo
  {publisher} {Butterworth-Heinemann},\ \bibinfo {year} {1987})\BibitemShut
  {NoStop}%
\bibitem [{\citenamefont {Chatterjee}\ \emph {et~al.}(2012)\citenamefont
  {Chatterjee}, \citenamefont {Parikh},\ and\ \citenamefont
  {Sarkar}}]{Chatterjee:2010gp}%
  \BibitemOpen
  \bibfield  {author} {\bibinfo {author} {\bibfnamefont {S.}~\bibnamefont
  {Chatterjee}}, \bibinfo {author} {\bibfnamefont {M.}~\bibnamefont {Parikh}},
  \ and\ \bibinfo {author} {\bibfnamefont {S.}~\bibnamefont {Sarkar}},\ }\href
  {\doibase 10.1088/0264-9381/29/3/035014} {\bibfield  {journal} {\bibinfo
  {journal} {Class. Quant. Grav.}\ }\textbf {\bibinfo {volume} {29}},\ \bibinfo
  {pages} {035014} (\bibinfo {year} {2012})},\ \Eprint
  {http://arxiv.org/abs/1012.6040} {arXiv:1012.6040 [hep-th]} \BibitemShut
  {NoStop}%
\bibitem [{Note3()}]{Note3}%
  \BibitemOpen
  \bibinfo {note} {For general $D \geq 4$, GB action is given by $S_{GB}=
  \DOTSI \intop \ilimits@ _M {\protect \mathcal L}_m \pm \DOTSI \intop
  \ilimits@ _{\partial M} {\protect \mathcal Q}_m$, where {\protect \it m} is
  the greatest integer $\leq D/2$ and ${\protect \mathcal L}_m = \Omega ^{a_1
  b_1} \wedge ... \wedge \Omega ^{a_m b_m} \wedge \Sigma _{a_1 b_1 ... a_m b_m}
  $ and ${\protect \mathcal Q}_m = m \DOTSI \intop \ilimits@ _0^1 \protect
  \text {ds} ~ \theta ^{a_1 b_1} \wedge \Omega _s^{a_2 b_2} \wedge ... \wedge
  \Omega _s^{a_m b_m} \wedge \Sigma _{a_1 b_1 ... a_m b_m}$. Here, $\Sigma
  _{a_1 b_1 \protect \dots a_m b_m}= \protect \frac {1}{(D-2m)!} \epsilon
  _{a_1b_1 \protect \dots a_m b_m c_1\protect \dots c_{D-2m}} e^{c_1} \wedge
  \protect \dots \wedge e^{c_{D-2m}}$ and $\Omega _s$ is the curvature of the
  connection $\omega _s=\omega - s \theta $, $\Omega _s = d\omega _s + \omega
  _s \wedge \omega _s$. The sign of the surface term depends upon the timelike
  or spacelike nature of the boundary. The surface term was first constructed
  in Ref. \cite {Myers:1987yn} and more details can be found
  there.}\BibitemShut {Stop}%
\bibitem [{\citenamefont {Davis}(2003)}]{Davis:2002gn}%
  \BibitemOpen
  \bibfield  {author} {\bibinfo {author} {\bibfnamefont {S.~C.}\ \bibnamefont
  {Davis}},\ }\href {\doibase 10.1103/PhysRevD.67.024030} {\bibfield  {journal}
  {\bibinfo  {journal} {Phys. Rev.}\ }\textbf {\bibinfo {volume} {D67}},\
  \bibinfo {pages} {024030} (\bibinfo {year} {2003})},\ \Eprint
  {http://arxiv.org/abs/hep-th/0208205} {arXiv:hep-th/0208205 [hep-th]}
  \BibitemShut {NoStop}%
\bibitem [{\citenamefont {Jacobson}\ and\ \citenamefont
  {Parentani}(2003)}]{Jacobson:2003wv}%
  \BibitemOpen
  \bibfield  {author} {\bibinfo {author} {\bibfnamefont {T.}~\bibnamefont
  {Jacobson}}\ and\ \bibinfo {author} {\bibfnamefont {R.}~\bibnamefont
  {Parentani}},\ }\href {\doibase 10.1023/A:1023785123428} {\bibfield
  {journal} {\bibinfo  {journal} {Found. Phys.}\ }\textbf {\bibinfo {volume}
  {33}},\ \bibinfo {pages} {323} (\bibinfo {year} {2003})},\ \Eprint
  {http://arxiv.org/abs/gr-qc/0302099} {arXiv:gr-qc/0302099 [gr-qc]}
  \BibitemShut {NoStop}%
\bibitem [{\citenamefont {Birmingham}(1999)}]{Birmingham:1998nr}%
  \BibitemOpen
  \bibfield  {author} {\bibinfo {author} {\bibfnamefont {D.}~\bibnamefont
  {Birmingham}},\ }\href {\doibase 10.1088/0264-9381/16/4/009} {\bibfield
  {journal} {\bibinfo  {journal} {Class. Quant. Grav.}\ }\textbf {\bibinfo
  {volume} {16}},\ \bibinfo {pages} {1197} (\bibinfo {year} {1999})},\ \Eprint
  {http://arxiv.org/abs/hep-th/9808032} {arXiv:hep-th/9808032 [hep-th]}
  \BibitemShut {NoStop}%
\bibitem [{\citenamefont {Brustein}\ and\ \citenamefont
  {Medved}(2009)}]{Brustein:2008cg}%
  \BibitemOpen
  \bibfield  {author} {\bibinfo {author} {\bibfnamefont {R.}~\bibnamefont
  {Brustein}}\ and\ \bibinfo {author} {\bibfnamefont {A.~J.~M.}\ \bibnamefont
  {Medved}},\ }\href {\doibase 10.1103/PhysRevD.79.021901} {\bibfield
  {journal} {\bibinfo  {journal} {Phys. Rev.}\ }\textbf {\bibinfo {volume}
  {D79}},\ \bibinfo {pages} {021901} (\bibinfo {year} {2009})},\ \Eprint
  {http://arxiv.org/abs/0808.3498} {arXiv:0808.3498 [hep-th]} \BibitemShut
  {NoStop}%
\bibitem [{\citenamefont {Cai}(2002)}]{Cai:2001dz}%
  \BibitemOpen
  \bibfield  {author} {\bibinfo {author} {\bibfnamefont {R.-G.}\ \bibnamefont
  {Cai}},\ }\href {\doibase 10.1103/PhysRevD.65.084014} {\bibfield  {journal}
  {\bibinfo  {journal} {Phys. Rev.}\ }\textbf {\bibinfo {volume} {D65}},\
  \bibinfo {pages} {084014} (\bibinfo {year} {2002})},\ \Eprint
  {http://arxiv.org/abs/hep-th/0109133} {arXiv:hep-th/0109133 [hep-th]}
  \BibitemShut {NoStop}%
\bibitem [{Note4()}]{Note4}%
  \BibitemOpen
  \bibinfo {note} {Our notation is different from the one in \cite
  {Brigante:2007nu}. We define the entropy density as the entropy per unit area
  of the horizon cross-section. For planar black holes the total entropy is
  infinite but the entropy density is well defined.}\BibitemShut {Stop}%
\bibitem [{\citenamefont {Cai}\ and\ \citenamefont {Soh}(1999)}]{Cai:1998vy}%
  \BibitemOpen
  \bibfield  {author} {\bibinfo {author} {\bibfnamefont {R.-G.}\ \bibnamefont
  {Cai}}\ and\ \bibinfo {author} {\bibfnamefont {K.-S.}\ \bibnamefont {Soh}},\
  }\href {\doibase 10.1103/PhysRevD.59.044013} {\bibfield  {journal} {\bibinfo
  {journal} {Phys. Rev.}\ }\textbf {\bibinfo {volume} {D59}},\ \bibinfo {pages}
  {044013} (\bibinfo {year} {1999})},\ \Eprint
  {http://arxiv.org/abs/gr-qc/9808067} {arXiv:gr-qc/9808067 [gr-qc]}
  \BibitemShut {NoStop}%
\bibitem [{\citenamefont {Aros}\ \emph {et~al.}(2001)\citenamefont {Aros},
  \citenamefont {Troncoso},\ and\ \citenamefont {Zanelli}}]{Aros:2000ij}%
  \BibitemOpen
  \bibfield  {author} {\bibinfo {author} {\bibfnamefont {R.}~\bibnamefont
  {Aros}}, \bibinfo {author} {\bibfnamefont {R.}~\bibnamefont {Troncoso}}, \
  and\ \bibinfo {author} {\bibfnamefont {J.}~\bibnamefont {Zanelli}},\ }\href
  {\doibase 10.1103/PhysRevD.63.084015} {\bibfield  {journal} {\bibinfo
  {journal} {Phys. Rev.}\ }\textbf {\bibinfo {volume} {D63}},\ \bibinfo {pages}
  {084015} (\bibinfo {year} {2001})},\ \Eprint
  {http://arxiv.org/abs/hep-th/0011097} {arXiv:hep-th/0011097 [hep-th]}
  \BibitemShut {NoStop}%
\bibitem [{\citenamefont {Banados}\ \emph {et~al.}(1994)\citenamefont
  {Banados}, \citenamefont {Teitelboim},\ and\ \citenamefont
  {Zanelli}}]{Banados:1993ur}%
  \BibitemOpen
  \bibfield  {author} {\bibinfo {author} {\bibfnamefont {M.}~\bibnamefont
  {Banados}}, \bibinfo {author} {\bibfnamefont {C.}~\bibnamefont {Teitelboim}},
  \ and\ \bibinfo {author} {\bibfnamefont {J.}~\bibnamefont {Zanelli}},\ }\href
  {\doibase 10.1103/PhysRevD.49.975} {\bibfield  {journal} {\bibinfo  {journal}
  {Phys. Rev.}\ }\textbf {\bibinfo {volume} {D49}},\ \bibinfo {pages} {975}
  (\bibinfo {year} {1994})},\ \Eprint {http://arxiv.org/abs/gr-qc/9307033}
  {arXiv:gr-qc/9307033 [gr-qc]} \BibitemShut {NoStop}%
\bibitem [{\citenamefont {Crisostomo}\ \emph {et~al.}(2000)\citenamefont
  {Crisostomo}, \citenamefont {Troncoso},\ and\ \citenamefont
  {Zanelli}}]{Crisostomo:2000bb}%
  \BibitemOpen
  \bibfield  {author} {\bibinfo {author} {\bibfnamefont {J.}~\bibnamefont
  {Crisostomo}}, \bibinfo {author} {\bibfnamefont {R.}~\bibnamefont
  {Troncoso}}, \ and\ \bibinfo {author} {\bibfnamefont {J.}~\bibnamefont
  {Zanelli}},\ }\href {\doibase 10.1103/PhysRevD.62.084013} {\bibfield
  {journal} {\bibinfo  {journal} {Phys. Rev.}\ }\textbf {\bibinfo {volume}
  {D62}},\ \bibinfo {pages} {084013} (\bibinfo {year} {2000})},\ \Eprint
  {http://arxiv.org/abs/hep-th/0003271} {arXiv:hep-th/0003271 [hep-th]}
  \BibitemShut {NoStop}%
\bibitem [{\citenamefont {Kolekar}\ and\ \citenamefont
  {Kothawala}(2012)}]{Kolekar:2011gg}%
  \BibitemOpen
  \bibfield  {author} {\bibinfo {author} {\bibfnamefont {S.}~\bibnamefont
  {Kolekar}}\ and\ \bibinfo {author} {\bibfnamefont {D.}~\bibnamefont
  {Kothawala}},\ }\href {\doibase 10.1007/JHEP02(2012)006} {\bibfield
  {journal} {\bibinfo  {journal} {JHEP}\ }\textbf {\bibinfo {volume} {02}},\
  \bibinfo {pages} {006} (\bibinfo {year} {2012})},\ \Eprint
  {http://arxiv.org/abs/1111.1242} {arXiv:1111.1242 [gr-qc]} \BibitemShut
  {NoStop}%
\bibitem [{\citenamefont {{V. E. Hubeny, S. Minwalla, and M. Rangamani,
  }}(2012)}]{Hubeny:2011hd}%
  \BibitemOpen
  \bibfield  {author} {\bibinfo {author} {\bibnamefont {{V. E. Hubeny, S.
  Minwalla, and M. Rangamani, }}},\ }\enquote {\bibinfo {title} {{The
  fluid/gravity correspondence}},}\ in\ \href@noop {} {\emph {\bibinfo
  {booktitle} {{Black Holes in Higher Dimensions}}}},\ \bibinfo {editor}
  {edited by\ \bibinfo {editor} {\bibnamefont {{Gary T. Horowtiz}}}}\ (\bibinfo
   {publisher} {Cambridge University Press},\ \bibinfo {year}
  {2012})\BibitemShut {NoStop}%
\bibitem [{\citenamefont {Eling}\ and\ \citenamefont
  {Oz}(2011)}]{Eling:2011ms}%
  \BibitemOpen
  \bibfield  {author} {\bibinfo {author} {\bibfnamefont {C.}~\bibnamefont
  {Eling}}\ and\ \bibinfo {author} {\bibfnamefont {Y.}~\bibnamefont {Oz}},\
  }\href {\doibase 10.1007/JHEP06(2011)007} {\bibfield  {journal} {\bibinfo
  {journal} {JHEP}\ }\textbf {\bibinfo {volume} {06}},\ \bibinfo {pages} {007}
  (\bibinfo {year} {2011})},\ \Eprint {http://arxiv.org/abs/1103.1657}
  {arXiv:1103.1657 [hep-th]} \BibitemShut {NoStop}%
\bibitem [{\citenamefont {Eling}\ \emph {et~al.}(2009)\citenamefont {Eling},
  \citenamefont {Fouxon},\ and\ \citenamefont {Oz}}]{Eling:2009pb}%
  \BibitemOpen
  \bibfield  {author} {\bibinfo {author} {\bibfnamefont {C.}~\bibnamefont
  {Eling}}, \bibinfo {author} {\bibfnamefont {I.}~\bibnamefont {Fouxon}}, \
  and\ \bibinfo {author} {\bibfnamefont {Y.}~\bibnamefont {Oz}},\ }\href
  {\doibase 10.1016/j.physletb.2009.09.028} {\bibfield  {journal} {\bibinfo
  {journal} {Phys. Lett.}\ }\textbf {\bibinfo {volume} {B680}},\ \bibinfo
  {pages} {496} (\bibinfo {year} {2009})},\ \Eprint
  {http://arxiv.org/abs/0905.3638} {arXiv:0905.3638 [hep-th]} \BibitemShut
  {NoStop}%
\bibitem [{\citenamefont {Buchel}(2011)}]{Buchel:2011yv}%
  \BibitemOpen
  \bibfield  {author} {\bibinfo {author} {\bibfnamefont {A.}~\bibnamefont
  {Buchel}},\ }\href {\doibase 10.1007/JHEP05(2011)065} {\bibfield  {journal}
  {\bibinfo  {journal} {JHEP}\ }\textbf {\bibinfo {volume} {05}},\ \bibinfo
  {pages} {065} (\bibinfo {year} {2011})},\ \Eprint
  {http://arxiv.org/abs/1103.3733} {arXiv:1103.3733 [hep-th]} \BibitemShut
  {NoStop}%
\bibitem [{\citenamefont {Faulkner}\ \emph {et~al.}(2011)\citenamefont
  {Faulkner}, \citenamefont {Liu},\ and\ \citenamefont
  {Rangamani}}]{Faulkner:2010jy}%
  \BibitemOpen
  \bibfield  {author} {\bibinfo {author} {\bibfnamefont {T.}~\bibnamefont
  {Faulkner}}, \bibinfo {author} {\bibfnamefont {H.}~\bibnamefont {Liu}}, \
  and\ \bibinfo {author} {\bibfnamefont {M.}~\bibnamefont {Rangamani}},\ }\href
  {\doibase 10.1007/JHEP08(2011)051} {\bibfield  {journal} {\bibinfo  {journal}
  {JHEP}\ }\textbf {\bibinfo {volume} {08}},\ \bibinfo {pages} {051} (\bibinfo
  {year} {2011})},\ \Eprint {http://arxiv.org/abs/1010.4036} {arXiv:1010.4036
  [hep-th]} \BibitemShut {NoStop}%
\bibitem [{\citenamefont {Donos}\ \emph {et~al.}(2017)\citenamefont {Donos},
  \citenamefont {Gauntlett}, \citenamefont {Griffin},\ and\ \citenamefont
  {Melgar}}]{Donos:2017oym}%
  \BibitemOpen
  \bibfield  {author} {\bibinfo {author} {\bibfnamefont {A.}~\bibnamefont
  {Donos}}, \bibinfo {author} {\bibfnamefont {J.~P.}\ \bibnamefont
  {Gauntlett}}, \bibinfo {author} {\bibfnamefont {T.}~\bibnamefont {Griffin}},
  \ and\ \bibinfo {author} {\bibfnamefont {L.}~\bibnamefont {Melgar}},\
  }\href@noop {} {\  (\bibinfo {year} {2017})},\ \Eprint
  {http://arxiv.org/abs/1701.01389} {arXiv:1701.01389 [hep-th]} \BibitemShut
  {NoStop}%
\end{thebibliography}%

\end{document}